\documentclass[lettersize,journal]{IEEEtran}
\usepackage{amsmath,amsfonts}
\usepackage{algorithmic}
\usepackage{algorithm}
\usepackage[caption=false,font=normalsize,labelfont=sf,textfont=sf]{subfig}
\usepackage{graphicx}
\usepackage{cite}
\usepackage{amssymb}
\usepackage{amsmath}
\usepackage{multirow}
\usepackage{hhline}
\usepackage{booktabs}
\usepackage{makecell}
\usepackage{comment}
\usepackage{xcolor}
\usepackage{url}

\definecolor{darkgreen}{rgb}{0.0, 0.5, 0.0}  

\begin{document}
\title{SCRec: A Scalable Computational Storage System with Statistical Sharding and Tensor-train Decomposition for Recommendation Models}

\author{Jinho Yang,~\IEEEmembership{Graduate Student Member,~IEEE}, Ji-Hoon Kim,~\IEEEmembership{Graduate Student Member,~IEEE},\\Joo-Young Kim,~\IEEEmembership{Senior Member,~IEEE,}

\thanks{This work was supported by Samsung Electronics Co., Ltd..}
\thanks{Manuscript received MM dd, yyyy; revised MM dd, yyyy.}}

\markboth{Journal of \LaTeX\ Class Files,~Vol.~nn, No.~nn, MM~yyyy}%
{Shell \MakeLowercase{\textit{et al.}}: A Sample Article Using IEEEtran.cls for IEEE Journals}


\maketitle

\begin{abstract}
Deep Learning Recommendation Models (DLRMs) play a crucial role in delivering personalized content across web applications such as social networking and video streaming. However, with improvements in performance, the parameter size of DLRMs has grown to terabyte (TB) scales, accompanied by memory bandwidth demands exceeding TB/s levels. Furthermore, the workload intensity within the model varies based on the target mechanism, making it difficult to build an optimized recommendation system. In this paper, we propose SCRec, a scalable computational storage recommendation system that can handle TB-scale industrial DLRMs while guaranteeing high bandwidth requirements. SCRec utilizes a software framework that features a mixed-integer programming (MIP)-based cost model, efficiently fetching data based on data access patterns and adaptively configuring memory-centric and compute-centric cores. Additionally, SCRec integrates hardware acceleration cores to enhance DLRM computations, particularly allowing for the high-performance reconstruction of approximated embedding vectors from extremely compressed tensor-train (TT) format. By combining its software framework and hardware accelerators, while eliminating data communication overhead by being implemented on a single server, SCRec achieves substantial improvements in DLRM inference performance. It delivers up to 55.77× speedup compared to a CPU-DRAM system with no loss in accuracy and up to 13.35× energy efficiency gains over a multi-GPU system.
\end{abstract}


\section{Introduction}

\begin{figure}[t]
    \centering
    \includegraphics[width=3.4in]{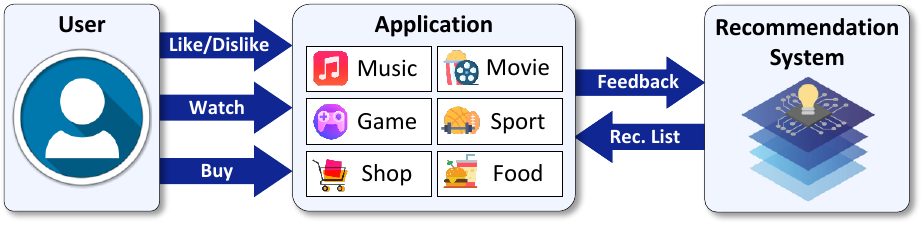}
    \caption{Overview of recommendation system.}
    \label{fig:overview_rm}
\end{figure}

\IEEEPARstart{R}{Recommendation} systems are widely used in social network services and video streaming platforms to provide personalized and preferred content to consumers as described in Fig.\ref{fig:overview_rm}. They are also employed in search engines to offer differentiated search services\cite{resnick1997recommender, raghuwanshi2019recommendation, darban2022ghrs, covington2016deep, cheng2016wide}. Supporting recommendation services for such a large number of consumers demands significant computation resources. For example, more than 80\% of Meta’s data center resources are allocated to recommendation system inference, while over 50\% are utilized for training these systems\cite{naumov2020deep}.

Traditional recommendation systems relied on collaborative filtering techniques, such as content filtering using matrix factorization\cite{goldberg1992using, hu2008collaborative, koren2009matrix, sarwar2001item}. However, with advancements in deep neural networks (DNNs), deep learning recommendation models (DLRMs) that combine embedding tables (EMBs) and multi-layer perceptron (MLP) layers have become a dominant approach in recommendation systems. These models are widely adopted in data centers, with recent focuses on both software-level and hardware-level optimizations\cite{naumov2019deep, shi2020compositional, gupta2020architectural, firoozshahian2023mtia, agarwal2023bagpipe, kwon2019tensordimm, gupta2021training}. DLRMs process categorical data like website visit histories through EMBs, and numerical data such as user age through MLP layers. This combination has demonstrated superior recommendation performance, making DLRM the industry standard in recommendation systems. However, their model size and memory bandwidth requirements have grown rapidly with the recommendation performance. Specifically, the number of EMBs, which accounts for the majority of a DLRM’s size, has increased to multiple terabytes (TBs) of parameters in memory devices, with bandwidth requirement rising to TB/s level\cite{zhao2020distributed, sethi2022recshard, zhao2019aibox}. 

In addition to the inherent characteristics of the DLRM algorithm, the workload intensity within the model varies depending on how it is applied in the recommendation system, making it difficult to build the optimized system\cite{gupta2020deeprecsys}. There are two primary mechanisms in recommendation systems: content filtering and content ranking. The former is the process of selecting a list of recommended content for users where embedding layer processing dominates, resulting in a memory-centric workload. In contrast, the latter determines the order in which the recommended content is displayed to users where MLP layer processing dominates, leading to a compute-centric workload. As a result, it is crucial to configure optimized recommendation systems with focusing on the targeting mechanism of the DLRM, to achieve high utilization of the system resources. Since traditional CPU-DRAM systems have struggled to meet the demands of modern DLRMs, it highlights the need for systems that can adaptively handle extensive memory and computational requirements.

Research has been conducted to address the memory and computational challenges of DLRM using solid-state drives (SSDs). Recent studies \cite{sun2022rm, kim2020reducing, wilkening2021recssd, wan2021flashembedding} have explored in-storage processing (ISP) and near-memory processing (NMP) to perform embedding layer computations within standalone SSDs. This approach facilitates the storage of TBs of EMBs and accelerates embedding layer operations by utilizing the internal memory bandwidth of SSDs, providing significantly faster performance compared to conventional SSD usage. However, the internal bandwidth of SSDs remains substantially lower than that of volatile memory, such as high-bandwidth memory (HBM), rendering them insufficient to meet the growing memory bandwidth demands of DLRMs. Furthermore, the high computational complexity of the MLP layer poses significant challenges for SSD-based ISP, making it impractical to implement a complete DLRM system.

Another approach to addressing both memory capacity and bandwidth is the use of multi-GPU systems. Recent works\cite{wang2022merlin, xiao2023g, mudigere2022software, wang2024rap} have implemented data parallelism and model parallelism to distribute the MLP and EMBs of the DLRM across multiple GPUs. Although this approach achieves highly performant systems, it also comes with a few challenging issues. The first issue is the large number of GPUs required to run a single DLRM. For instance, constructing a DLRM system with TBs of parameters would require dozens of NVIDIA H100 GPUs, each equipped with 80 GB of HBM, resulting in enormous system costs. The second issue is the underutilization of GPU compute resources for embedding-dominant DLRM applications, leading to inefficiency in the entire system. Lastly, multiple GPU nodes must be interconnected to run the entire DLRM system at TB scale in parallel, leading to increased network complexity and overhead. The overall DLRM performance degradation is inevitable due to complex network configuration among GPU nodes, with data communication overhead among them. To be precise, all-to-all communication is required during model inference, where each GPU requests embedding vectors from different GPUs, causing significant system bottlenecks.

To mitigate these challenges, recent work\cite{sethi2022recshard} has statistically analyzed DLRM characteristics and proposed a sharding technique that stores frequently accessed embedding vectors in the GPU’s HBM, while placing rarely accessed embedding vectors in the host’s DRAM. This approach effectively maintains a comparable system memory bandwidth while reducing the required number of GPUs to run the DLRM. However, challenges such as high system cost, computing resource underutilization, and complexity of network topology still persist in the GPU-based DLRM systems.

To address these challenges, we propose SCRec, a scalable computational storage system that can handle TB-scale industrial DLRMs while providing high bandwidth requirements. SCRec leverages SmartSSD devices, which integrate an SSD with a field-programmable gate array (FPGA) for near-data processing (NDP), effectively utilizing the SSD’s large capacity alongside the high bandwidth of DRAM and block RAM (BRAM) on the FPGA. The detailed contributions of our work are as follows.

\begin{itemize}
\item
SCRec utilizes a software framework with two primary features—statistical sharding and adaptive acceleration core mapping—leveraging a mixed-integer programming (MIP)-based cost model. By analyzing data access patterns of DLRM input features, SCRec optimizes SmartSSD memory bandwidth by placing hot data in high-bandwidth memory devices such as BRAM and DRAM, and cold data in SSD. Additionally, it determines the optimal configuration of memory-centric and compute-centric acceleration cores based on the workload intensity analysis for each layer, ensuring a high-utilization DLRM system.
\item
We designed custom acceleration cores capable of boosting up EMB and MLP computations in DLRM and implemented them onto the FPGA chip of the SmartSSD device. Especially, the EMB core utilizes the tensor-train (TT) format to compress the GB-sized EMBs significantly into the megabyte (MB)-level, enabling the high-performance reconstruction of approximated embedding vectors. This also complements the device's DRAM capacity limitations while enhancing the overall system’s memory bandwidth.
\item
We integrated the software framework and hardware accelerators to develop a multi-SmartSSD system capable of running the entire DLRM operations. The system is configured within a single server, eliminating the need for complex network setup and eliminating data communication overhead. Additionally, within the server, it minimizes host communication overhead through peer-to-peer (P2P) data transfer between the SSD and FPGA chip. Therefore, SCRec can significantly reduce both inter-node and intra-node communication overhead.

\end{itemize}

Our evaluation demonstrates that SCRec achieves significant improvements in inference performance, achieving up to 55.77$\times$ speed-up compared to a CPU-DRAM system with no loss in accuracy and up to 13.35$\times$ improvement in energy efficiency over a multi-GPU system.

\begin{figure}[t]
    \centering
    \includegraphics[width=3.4in]{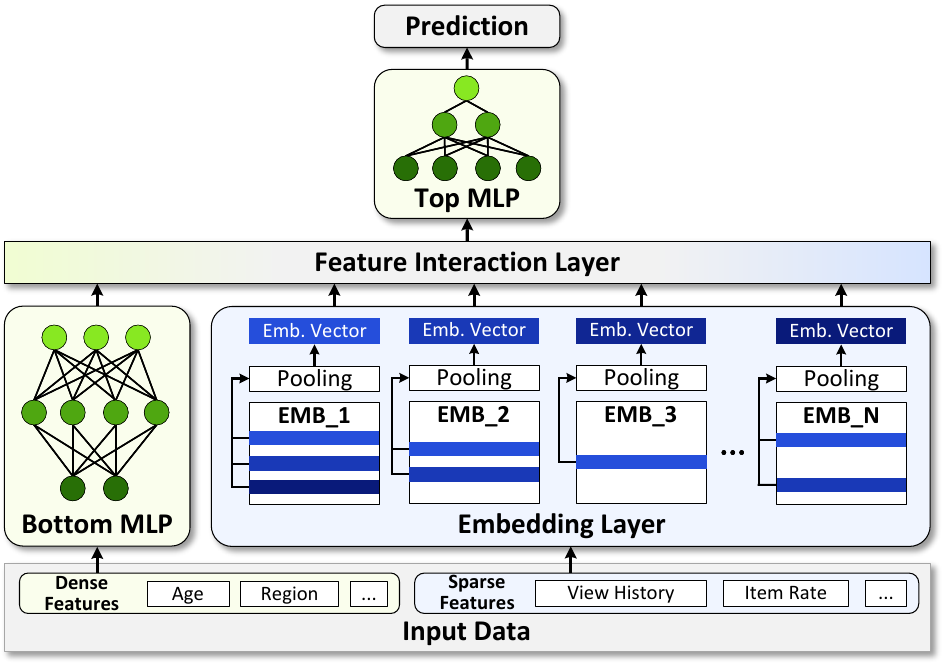}
    \caption{Architecture of deep learning recommendation model.}
    \label{fig:dlrm_arch}
\end{figure}

The remainder of this paper is organized as follows: Section \ref{sec:back_gnd} provides background information on the DLRM and TT-format; Section \ref{sec:impl_sys} describes SCRec, including its software framework and hardware architecture; Section \ref{sec:eval} presents the evaluation results of SCRec; Section \ref{sec:rel_work} discusses related works on memory caching, sharding, and TT decomposition; and Section \ref{sec:conc} concludes the paper.

\section{Background}
\label{sec:back_gnd}

\subsection{DLRM Architecture and Hybird-parallel Processing}

DLRM is a DNN-based recommendation model widely utilized across various industries, as depicted in Fig.\ref{fig:dlrm_arch}. It processes user and content data as input to predict user interactions with specific content, aiming to estimate the click-through rate (CTR). The model primarily consists of an embedding layer with multiple EMBs, MLP layers, and a feature interaction layer. In the DLRM, dense features and sparse features serve as inputs to the MLP layer and the embedding layer, respectively. Dense features, also referred to as continuous features, represent numerical data such as a user’s age or geographic location. Sparse features, also known as categorical features, include data like website visit history or content category IDs. Sparse features are specifically represented as one-hot or multi-hot binary vectors, which are used to retrieve latent vectors from the EMBs. These latent vectors are then aggregated through a pooling operation to produce a single vector. The outputs of the bottom MLP layer and the embedding layer are integrated through concatenation in the feature interaction layer, followed by computation in the top MLP layer to produce the final prediction. A key strength of DLRM is its ability to use multiple EMBs to map discrete, categorical features into their corresponding vector representations, allowing it to effectively capture semantically meaningful representations of the target features\cite{kwon2021tensor}.

\begin{figure}[t]
    \centering
    \includegraphics[width=3.4in]{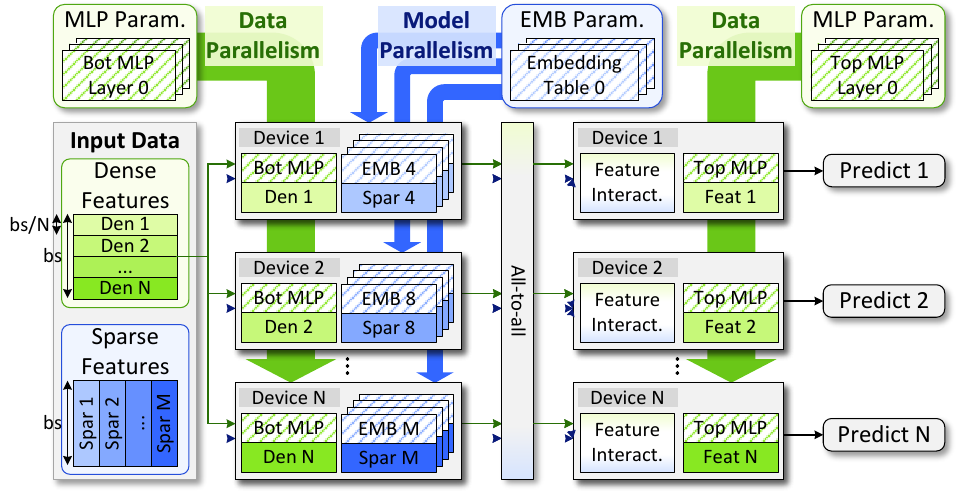}
    \caption{Hybrid-parallel DLRM processing.}
    \label{fig:dlrm_parallel}
\end{figure}

The need for parallel processing across multiple devices in the DLRM arises from the model's inherent characteristics. First, EMBs, which account for the majority of DLRM parameters, can number hundreds, with the total size of the embedding layer exceeding several TBs. Consequently, memory devices utilizing DRAM and HBM must split and store the EMBs across multiple devices. Second, as the MLP layer is compute-bound and the embedding layer is memory-bound, optimizing the system with respect to workload intensity is essential to prevent hardware underutilization. To address these challenges, hybrid-parallel processing is applied to optimize DLRM execution across multiple devices, as depicted in Fig.\ref{fig:dlrm_parallel}. Data parallelism is used for the MLP layer, whereas model parallelism is applied to the embedding layer. Specifically, dense features are divided among devices using a batch-wise split, and EMBs are configured with a table-wise split. In this setup, an all-to-all operation ensures that each device holds all pooled latent vectors, as the feature interaction layer requires pooling results from all EMBs. Subsequently, the feature interaction layer and top MLP layer are processed using data parallelism, finally producing the CTR results. Hybrid-parallel processing in DLRM is a powerful approach that reduces the computational intensity of MLP operations while overcoming memory capacity limitations and increasing effective memory bandwidth.

\subsection{Tensor-train Decomposition}

TT decomposition\cite{oseledets2011tensor} is a low-rank approximation method that decomposes high-dimensional tensors into lower-dimensional representations, enabling them to be approximated with significantly less data. The decomposition process involves the following steps: A d-dimensional tensor \(\mathcal{T}\in\mathbb{R}^{I_1 \times I_2 \times \dots \times I_d} \), where $I_k$ is the size of dimension $k$, can be represented as a sequential product of tensors \(\mathcal{G}\) with lower dimension d, using the following equation \ref{eq:tt_rep}.

\begin{figure}[t]
    \centering
    \includegraphics[width=3.4in]{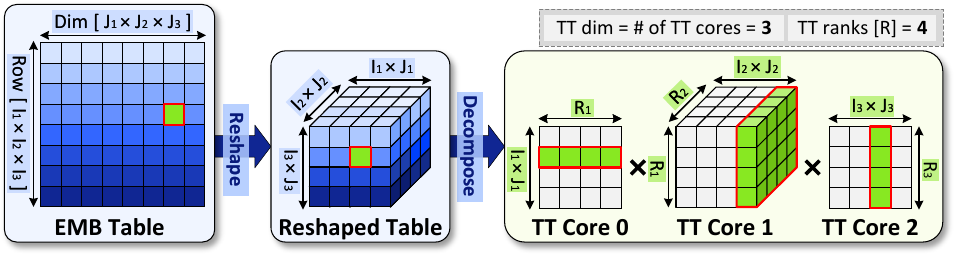}
    \caption{Process of representing an embedding table in TT-format.}
    \label{fig:ttrep}
\end{figure}

\begin{equation}
\mathcal{T}(i_1, i_2, \dots, i_d) = \mathcal{G}_1(:, i_1, :) \mathcal{G}_2(:, i_2, :) \dots \mathcal{G}_d(:, i_d, :)
\label{eq:tt_rep}
\end{equation}

The decomposed tensor \( \mathcal{G}_k \in \mathbb{R}^{R_{k-1} \times I_k \times R_k} \) is referred to as a TT-core, has a TT-rank \( R_k \), where the first and last TT-cores have TT-rank of 1 to ensure that the sequential product of TT-cores results in a scalar, meaning \( R_0 = R_k = 1 \).

A matrix-form EMB can be represented using TT-format, as illustrated in Fig.\ref{fig:ttrep}. If the EMB is represented as \(E \in \mathbb{R}^{I \times J} \) , where  \( I = \prod_{k=1}^{d} i_k \) refers to the row length and  \( J = \prod_{k=1}^{d} j_k \)  refers to embedding dimension. the EMB can be reshaped into a d-dimensional tensor  \( \mathcal{T}\in\mathbb{R}^{(i_1j_1) \times (i_2j_2) \times \dots \times (i_dj_d)}\). This reshaped tensor can then be decomposed into multiple TT-cores. Consequently, every element of the EMB can be approximated as a value obtained through the sequential product of TT-cores, which can be computed using Equation \ref{eq:tt_emb}:

\begin{equation}
\begin{aligned}
    E(i, j) &= \mathcal{T}(m_1, m_2, \dots, m_d) \\
        &= \mathcal{G}_1(:, m_1, :) \mathcal{G}_2(:, m_2, :) \dots \mathcal{G}_d(:, m_d, :)
\label{eq:tt_emb}
\end{aligned}
\end{equation}

where the EMB row is defined as \( i = \sum_{k=1}^{d} i_k \prod_{l=k+1}^{d} I_l \), the EMB dimension value by \( j = \sum_{k=1}^{d} j_k \prod_{l=k+1}^{d} J_l \), and the reshaped tensor index by  \( m_d = I_d \cdot i_d + j_d \), with the ranges \( 0 \leq i_d \leq I_d-1\) and \( 0 \leq j_d \leq J_d-1\)\cite{yin2021tt}.

\begin{figure*}[t]
    \centering
    \includegraphics[width=6.8in]{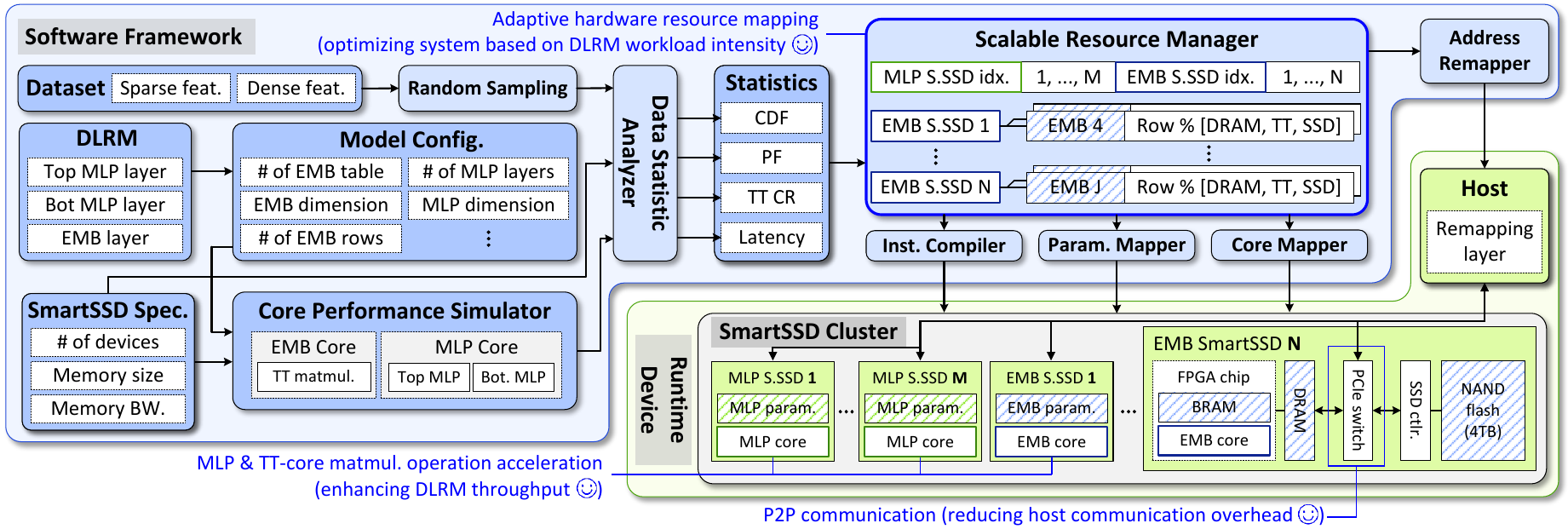}
    \caption{Overview of SCRec.}
    \label{fig:flexrec_sys}
\end{figure*}

The use of TT decomposition to compress DNN model parameters has significantly reduced the need for large memory capacity. This method has proven particularly effective, achieving exceptionally high compression ratios—up to thousands of times—for DLRMs, especially for EMBs, which range from hundreds of GBs to severals TBs. Additionally, fetching parameters in TT-format to on-chip memory reduces the off-chip memory footprint, thereby enhancing the model throughput.
\section{SCRec}
\label{sec:impl_sys}

\subsection{Overview of SCRec}

SCRec integrates a software framework that configures an optimized DLRM system based on DLRM’s workload intensity, along with a custom hardware accelerator to enhance model performance. SCRec consists of three main pipelines in the offline phase: the Data Statistic Analyzer (DSA), the scalable Resource Manager (SRM), and the Address Remapper. Detailed descriptions of each stage are provided as follows:

\textbf{Data Statistic Analyzer}
DSA identifies the statistical characteristics of a dataset and uses them as parameters for the cost model in the subsequent SRM task, providing the necessary information for analyzing workload. Specifically, the DSA utilizes a randomly subsampled input dataset, the configuration of the DLRM, and the hardware specifications of the SmartSSD to derive key metrics, as explained in Section \ref{subsec:soft_dsa}. SCRec simulates hardware acceleration logic mapped to the SmartSSD using a cycle-accurate performance simulator. This simulation measures the latency required to retrieve embedding vectors in an EMB core and the latency of MLP operations in an MLP core while considering the off-chip memory footprint. These metrics are also used as parameters to build the cost model in the SRM.

\textbf{Scalable Resource Manager}
Based on the statistics obtained from the DSA, the SRM applies a MIP-based cost model to enable adaptive hardware core allocation and model parameter mapping on the SmartSSDs, based on the workload intensity of DLRM. A key feature of the SRM is its three-level sharding strategy, which stores hot embedding vectors with high access frequency in FPGA DRAM and BRAM, while cold embedding vectors with low access frequency are stored in the SSD. This approach leverages the large memory capacity of SSDs and the high memory bandwidth of DRAM. Notably, by storing EMBs in TT-format within FPGA BRAM, even large EMBs can be accommodated in on-chip memory, effectively overcoming the capacity limitations of FPGA DRAM, as discussed in Section \ref{subsec:soft_srm}. Additionally, the SRM enables scalable, hardware-based latency optimizing by adaptively allocating MLP and EMB cores based on the workload intensity of the DLRM. Using the results generated by the SRM, heterogeneous computation cores are mapped to the SmartSSD, and DLRM parameters are fetched. Simultaneously, the instruction compiler generates instructions for multi-device execution.

\textbf{Addresss Remapper}
In the address remapper, a remapping layer is implemented to translate logical memory addresses, enabling access to sharded EMBs stored in corresponding devices. During DLRM execution, this remapping layer is loaded into host DRAM at runtime, facilitating memory access when the host requests embedding layer access from SmartSSDs. The implementation details of the remapping layer are discussed in Section \ref{subsec:soft_remap}.

\textbf{Hardware Accelerator}
On the hardware side, the MLP core and EMB core are designed to accelerate computations in the MLP layer and embedding layer, respectively. In particular, the primary function of the EMB core is to generate approximated embedding vectors represented in the TT-format while storing them on-chip, as detailed in Section \ref{subsec:core_arch}. Additionally, P2P communication in SmartSSD is implemented via a PCIe switch to directly connect the SSD and FPGA chip, minimizing host communication overhead through NDP.

\subsection{Data Statistic Analyzer}
\label{subsec:soft_dsa}

\begin{figure}[t]
    \centering
    \includegraphics[width=3.4in]{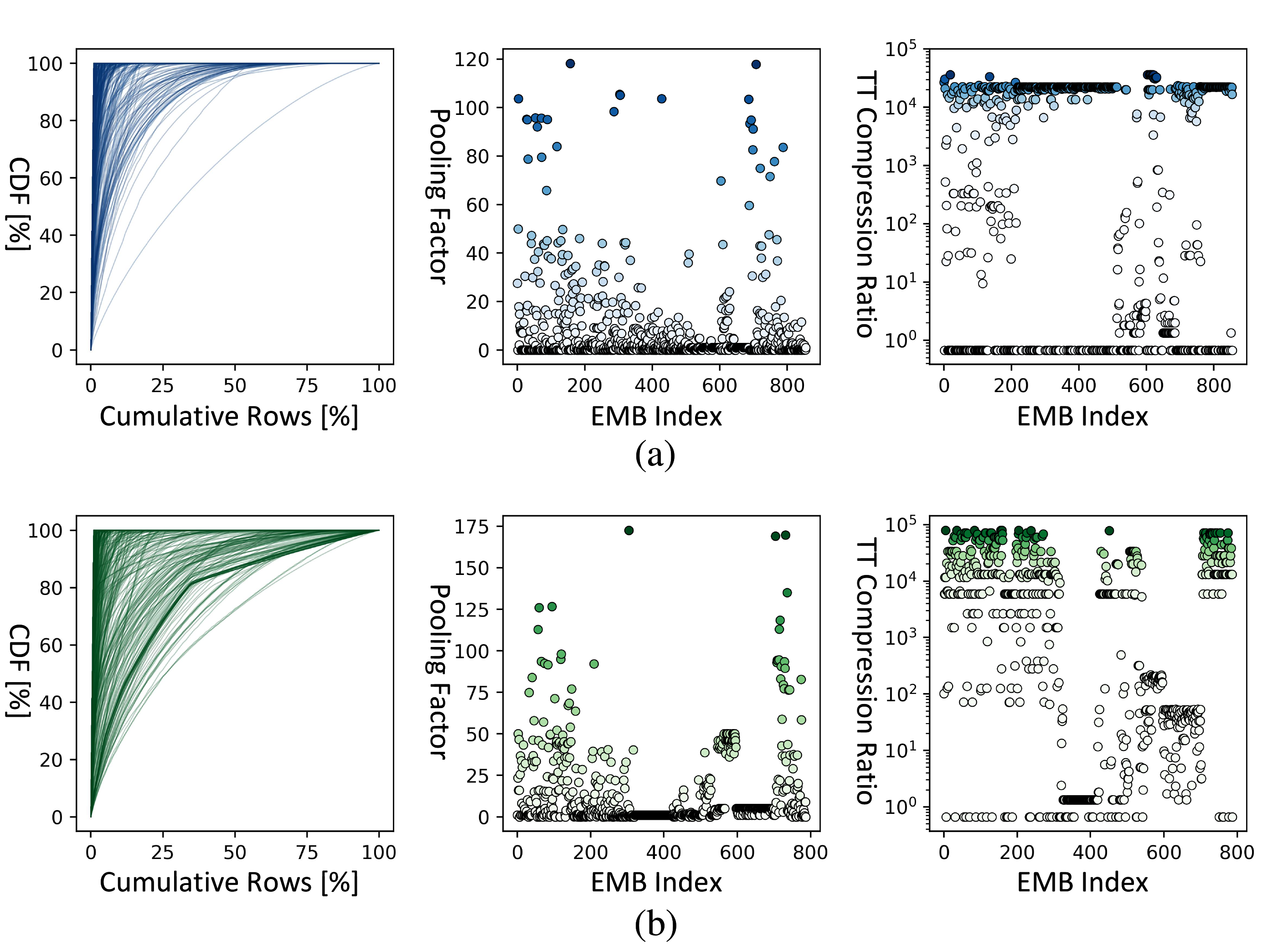}
    \caption{Characteristic of sparse featrues in Meta embedding lookup synthetic dataset. (a) Year 2021 (b) Year 2022}
    \label{fig:dlrm_char_graph}
\end{figure}

The goal of DSA is to derive the parameters required for the cost model used in the SRM. This is achieved using sub-sampled data, the model configuration, and a core performance simulator. The DSA computes essential parameters for constructing the cost model, including the cumulative distribution function (CDF), pooling factor (PF), TT compression ratio (CR), and layer operation latency. In SCRec, statistics for the Meta Embedding Lookup Synthetic (MELS) dataset \cite{github_meta_syn} from 2021 and 2022, which simulates the embedding layer of Meta’s industrial DLRM, were analyzed. The results of this data profiling are presented in Fig.\ref{fig:dlrm_char_graph}(a) and Fig.\ref{fig:dlrm_char_graph}(b), respectively. Detailed explanations of each statistical metric are provided below:

\textbf{Cumulative Distribution Function}
CDF is used to describe how frequently different rows in EMBs are accessed within the data sample, as shown in the left graph of Fig.\ref{fig:dlrm_char_graph}. Specifically, we analyzed the 856 EMBs from the 2021 dataset and the 788 EMBs from the 2022 dataset, plotting graphs of the access statistics. These graphs reveal that the EMB access pattern follows a flipped power-law distribution, where a small portion of the EMB accounts for the majority of total accesses. This observation suggests that employing a multi-level memory hierarchy—with hot data placed in smaller, high-bandwidth memory and cold data stored in larger, low-bandwidth memory—can significantly improve EMB access performance.

\textbf{Pooling Factor}
PF refers to the number of embedding vectors accessed for a particular sparse feature in a given data sample. PF directly impacts memory bandwidth demand, as a higher PF requires accessing more EMB rows, thereby increasing the memory load. The middle graphs in Fig.\ref{fig:dlrm_char_graph} illustrate the average PF, showing a wide range from 0 to 118 in the 2021 dataset and from 0 to 175 in the 2022 dataset. Due to the significant variance in PF across EMBs, PF becomes a critical factor in effectively partitioning EMBs across devices.

\textbf{TT Compression Ratio}
TT CR refers to the ratio that quantifies how much the data can be compressed when an EMB is represented in TT-format. It is expressed as $CR = EMB\_size / TT\_core\_size$. Calculating the CR in the DSA is crucial because TT-cores must fit within the limited capacity of BRAM, which is only a few MBs. Additionally, the CR depends on various factors, including the rank of the TT-core, the dimensions of the EMBs, and the percentage of rows in the EMB selected for compression. For example, the graph on the right side of Fig.\ref{fig:dlrm_char_graph} shows the CR when the entire EMB is compressed with an embedding dimension of 64 and a TT-rank of 4. In some EMBs, the TT-represented EMB surpasses the original size, while in others, the CR reaches values in the thousands. This variation arises because larger EMBs generally achieve higher CR. Given this characteristic, it is crucial to construct a cost model that incorporates the TT CR of each EMB and its resulting compressed size. This ensures that TT-format EMBs can be effectively selected and compressed to fit within the FPGA's on-chip memory in the SmartSSDs using the cost model, thereby optimizing memory usage and access.

\textbf{Layer Operation Latency}
The latency of generating approximated embedding vectors from TT-cores, along with the processing latency of the top and bottom MLP layers, is obtained through the core performance simulator and used as constraints in the cost model. Decoupling the cost model from the latency measurement simulator reduces the computational overhead of the cost model, while the cycle-accurate design of the simulator ensures precise latency estimations. Notably, when calculating the latency of MLP layers, the off-chip memory footprint is also considered, resulting in even more accurate latency measurements.

\begin{table}[t]
    \begin{center}
        \caption{Descriptions of SRM cost model parameters}
        \label{tab:fhrm_param}
        \begin{tabular}{cl}
            \toprule
            \textbf{Parameter} & \textbf{Description}   \\ \hhline{==}
            $M$ & Number of SmartSSDs   \\
            $J$ & Number of EMBs   \\
            $BS$ & Batch Size   \\
            $BS\_mini$ & Mini Batch Size of MLP Core   \\
            $Cap_{bram}$ & FPGA BRAM Capacity per SmartSSD   \\
            $Cap_{dram}$ & FPGA DRAM Capacity per SmartSSD   \\
            $Cap_{ssd}$ & SSD Capacity per SmartSSD   \\
            $t_{tt}$ & TT-core Matrix multiplication Latency   \\
            $t_{dram}$ & DRAM Read Latency   \\
            $t_{ssd}$ & SSD Read Latency   \\
            $t_{mlp\_top}$ & Top MLP Layer Processing Latency   \\
            $t_{mlp\_bot}$ & Bottom MLP Layer Processing Latency   \\
            $ICDF_j$ & Inverse CDF of EMB $j$   \\
            $step_j$ & Step Size for Piecewise Linear Interpolation of EMB $j$   \\
            $tt\_cm_j$ & TT-cores Size of EMB $j$   \\
            $pf_j$ & Average Pooling Factor of EMB $j$   \\
            $row\_len_j$ & Row Length of EMB $j$   \\
            $dim_j$ & Embedding Dimension of EMB $j$   \\
            $emb_j$ & Overall Data Size of EMB $j$   \\
            $df_j$ & Data Format Byte Size of EMB $j$   \\
            $hot\_thr_j$ & Hot Data Threshold Percentage of EMB $j$  \\
            \bottomrule
        \end{tabular}
    \end{center}
\end{table}

\subsection{Scalable Resource Manager}
\label{subsec:soft_srm}

SRM aims to achieve configuring optimized DLRM system through a MIP-based cost model, leveraging the characteristics of the DLRM profiled by the DSA. The core concept of SRM is to minimize EMB access latency by utilizing a three-level memory hierarchy and the TT-format. Specifically, EMBs requiring high bandwidth are fetched into the DRAM and BRAM of the FPGA, while those with lower bandwidth requirements are stored in the SSD. This strategy enables the embedding layer parameters to take advantage of the large capacity of SSDs while benefiting from the high bandwidth of DRAM and BRAM. Moreover, by storing EMBs approximated in the highly compressed TT-format within the BRAM, it becomes possible to fit GB-sized EMBs into the MB-sized on-chip memory, effectively increasing the embedding layer’s effective bandwidth. We refer to this technique as three-level sharding, and the following explanation details the implementation of the cost model used to apply this sharding scheme.

\textbf{Objective Function}
To optimize the latency of DLRM using multiple SmartSSD devices, we define an objective function to minimize the cost, as described in Equation \ref{eq:obj_func}, with the parameters outlined in Table \ref{tab:fhrm_param}.

\begin{equation}
    \begin{aligned}
        &\text{minimize} \quad C \\
        &\text{subject to} \quad c_{fnt} + c_{mlp\_top} \leq C
    \end{aligned}
    \label{eq:obj_func}
\end{equation}

$c_{fnt}$ refers to the cost of the layer preceding the feature interaction layer, while $c_{mlp\_top}$ represents the cost of the top MLP layer. Since the bottom MLP layer and the embedding layer, which together determine $c_{fnt}$, are processed in parallel, the cost function is designed to minimize the sum of the maximum cost between these two independently processed layers and the cost of the top MLP layer.

\textbf{Device Allocation}
The constraints for assigning cores capable of computing the MLP and embedding layers are as follows:

\begin{align}
    1 \leq \sum_{m} d_m \leq M - 1 \quad \forall m \in M \label{eq:dev_alloc} \\
    d_m \in \{0, 1\} \quad \forall m \in M
\end{align}

$d_m$ represents the type of hardware core to be mapped onto the FPGA chip of the SmartSSD. Specifically, when $d_m=0$, the constraint ensures that an MLP core is mapped, whereas when $d_m=1$, it enforces the mapping of an EMB core. Equation \ref{eq:dev_alloc} controls the allocation of heterogeneous cores, preventing all SmartSSDs from being exclusively assigned to either EMB cores or MLP cores. This allocation ensures that all layers of the DLRM can be effectively accelerated within the multi-SmartSSD system.

\textbf{Embedding Table Allocation}
To enable model parallelism by partitioning the EMB parameters across SmartSSDs mapped to EMB cores, the following constraints are defined:

\begin{align}
    \sum_{m} p_{mj} = 1 \quad \forall j &\in J \label{eq:p_mj_1} \\
    \sum_{j} p_{mj} = d_{m} * \sum_{j} p_{mj} \quad \forall m &\in M \label{eq:p_mj_2} \\
    p_{mj} \in \{0, 1\} \quad \forall m &\in M 
\end{align}

$p_{mj}$is a binary variable that indicates whether EMB $j$ is assigned to SmartSSD $m$. In Equation \ref{eq:p_mj_1}, a constraint is imposed to ensure that each specific EMB is assigned to its corresponding SmartSSD. Equation \ref{eq:p_mj_2} ensures that EMBs are only assigned to SmartSSDs that have been allocated EMB cores, serving as a constraint to prevent the incorrect assignment of EMBs to SmartSSDs with MLP cores. By utilizing these constraints, SCRec achieves model parallelism in the embedding layer through table-wise splitting of EMBs in the SRM.

\textbf{Three-level Sharding}
The EMBs assigned to SmartSSDs are split row-wise based on their access frequency patterns and are fetched into the FPGA’s DRAM, BRAM, and SSD memory devices. The constraints are structured as follows to minimize EMB access latency:

\begin{equation}
\begin{aligned}
    \sum_{i} &x\_dram_{ji} * ICDF_{j}(i) * row\_len_{j} \\[-7pt]
    \hspace{0.6cm} *& \hspace{0.1cm} dim_{j} * df_{j} = mem\_dram_{j} \quad \forall j \in J \label{eq:mem_dram_1}
\end{aligned}
\end{equation}

\vspace{-0.3cm}

\begin{align}
    \sum_{i} x\_dram_{ji} * \frac{i}{step_{j}} = pct\_dram_{j} \quad \forall j &\in J \label{eq:pct_dram_2}\\
    \sum_{i} x\_dram_{ji} = 1 \quad \forall j &\in J \label{eq:pct_dram_3} \\
    x\_dram_{ji} \in \{0, 1\} \quad i = 0, \dots, step_{j} \quad \forall j &\in J \label{eq:pct_dram_4}
\end{align}

The above four equations constrain which rows of EMB $j$ are fetched into the FPGA DRAM and determine the size of those rows. The variable $x\_dram$ serves as a binary variable to compute the $pct\_dram$ variable, which determines , which specifies the percentage of rows from EMB $j$ are stored in DRAM, as described in Equation \ref{eq:pct_dram_2}. At this point, the step size for EMB $j$ is set to min($row\_len_j$, 100). For the ICDF, the CDF obtained from the DSA is inverted and the interpolation interval is determined by $step_j$ using piecewise linear interpolation, reducing the computational complexity of the cost model. Equation \ref{eq:mem_dram_1}, formulated using the ICDF, calculates the size of EMB $j$ to be fetched into DRAM, with $mem\_dram_j$ serving as a constraint on the FPGA DRAM capacity of the SmartSSD.

\begin{equation}
\begin{split}
\sum_{i} x\_ptr\_tt_{ji} * ICDF_{j}(i) * row\_len_{j} * dim_{j} * df_{j} \\
- mem\_dram_{j} = mem\_tt_{j} \quad \forall j \in J \label{eq:mem_tt_1}
\end{split}
\end{equation}

\vspace{-0.3cm}

\begin{equation}
\begin{split}
\sum_{i} x\_dram_{ji} * \frac{i}{step_{j}} + \sum_{i} x\_tt_{ji} * \frac{i}{step_{j}} \\
\hspace{1.9cm}  = \sum_{i} x\_ptr\_tt_{ji} * \frac{i}{step_{j}} \quad \forall j \in J \label{eq:ptr_tt_2}
\end{split}
\end{equation}

\vspace{-0.5cm}

\begin{equation}
\hspace{1.3cm} \sum_{i} x\_tt_{ji} * \frac{i}{step_{j}} = pct\_tt_{j} \quad \forall j \in J \label{eq:pct_tt_3}\\
\end{equation}

\vspace{-0.5cm}

\begin{equation}
\quad \quad \quad \sum_{i} x\_tt_{ji} = \sum_{i} x\_ptr\_tt_{ji} = 1 \quad \forall j \in J
\end{equation}

\vspace{-0.5cm}

\begin{equation}
\hspace{0.9cm} x\_tt_{ji} \in \{0, 1\} \quad i = 0, \dots, step_{j} \quad \forall j \in J
\end{equation}

\vspace{-0.5cm}

\begin{equation}
\hspace{0.3cm} x\_ptr\_tt_{ji} \in \{0, 1\} \quad i = 0, \dots, step_{j} \quad \forall j \in J
\end{equation}

The above six equations define constraints for determining which rows of EMB $j$ are fetched into BRAM in the compressed TT-format and the size of those rows before compression. The variable $x\_tt$ is a binary variable used to calculate the percentage of rows from EMB $j$ stored in BRAM, as described in Equation \ref{eq:pct_tt_3}, while $x\_ptr\_tt$ is a binary variable used to calculate the size of the EMB rows allocated to BRAM. Unlike Equations \ref{eq:mem_dram_1}--\ref{eq:pct_dram_4}, the EMB rows stored in TT-format are positioned between the rows stored in DRAM and SSD. Thus, the starting and ending row indices of the EMB stored in TT-format can be derived using the $x\_dram$ and $x\_ptr\_tt$ binary variables via Equations \ref{eq:ptr_tt_2} and \ref{eq:pct_tt_3}. Additionally, the size of EMB $j$ before compression into TT-format, denoted as $mem\_tt_j$, is calculated using the ICDF in Equation \ref{eq:mem_tt_1}, and this variable serves as a constraint on the BRAM capacity of the SmartSSD.

\begin{equation}
\begin{aligned}
\sum_{i} x\_ptr\_tt_{ji} * ICDF_{j} - \sum_{i} x\_dram_{ji} * ICDF_{j} \\
= \sum_{i} x\_row\_tt_{ji} * \frac{i}{step_{j}} \quad \forall j \in J \\
\end{aligned}
\end{equation}

\vspace{-0.3cm}

\begin{equation}
\hspace{2.2cm} \sum_{i} x\_row\_tt_{ji} = 1 \quad \forall j \in J
\end{equation}

\vspace{-0.5cm}

\begin{equation}
\hspace{0.1cm} x\_row\_tt_{ji} \in \{0, 1\} \quad i = 0, \dots, step_{j}  \quad \forall j \in J
\end{equation}

The above three equations define constraints for calculating the $x\_row\_tt$ binary variable, which is necessary to determine the size of EMB $j$ when represented across multiple TT-cores. Using $x\_row\_tt$, the cost model calculate the size of the EMBs compressed into TT-format, enabling the formulation of BRAM capacity constraints.

\begin{equation}
   \hspace{1.1cm} pct\_dram_{j} + pct\_tt_{j} \leq hot\_thr_j \quad \forall j \in J \label{eq:hot_thr}
\end{equation}

Equation \ref{eq:hot_thr} imposes constraints to prevent excessive TT-format compression, ensuring that less frequently accessed embedding vectors are stored in the SSD. The $hot\_thr_{j}$ is assigned a value for each EMB. A value of 1 is used when the EMB row size is relatively small, whereas a value less than 1 is applied for significantly larger row sizes, ensuring that EMB parameters from larger tables are fetched from the SSD. 

\textbf{Memory Capacity}
The constraints in the EMB allocation process determine how each EMB $j$ is split at the row level and the extent of memory capacity it occupies when assigned to a SmartSSD. Additionally, a condition must be imposed to ensure that the size of the EMBs fetched to the device does not exceed the device's memory capacity. The constraints are defined as follows:

\begin{align}
\hspace{0.6cm} row\_len_{j} * dim_{j} * df_{j} = emb_{j} \quad \forall j &\in J \label{eq:emb_cap_1}\\
\hspace{0.6cm} \sum_{j} p_{mj} * mem\_dram_{j} \leq Cap_{DRAM} \quad \forall m &\in M \label{eq:emb_cap_2}
\end{align}

\vspace{-0.5cm}

\begin{equation}
\begin{split}
    \sum_{j} p_{mj} * (emb_{j} - mem\_dram_{j} - mem\_tt_{j}) \\[-7pt]
    \hspace{4.6cm} \leq Cap_{SSD} \quad \forall m \in M \label{eq:emb_cap_3}
\end{split}
\end{equation}

\vspace{-0.3cm}

\begin{align}
    \sum_{j} x\_row\_tt_{ij} * tt\_cm_{j}(i) * df_{j} = tt\_cap_{j} \quad \forall j &\in J \label{eq:emb_cap_4} \\[-7pt]
    \sum_{i} p_{mj} * tt\_cap_{j} \leq Cap_{BRAM} \quad \forall m &\in M \label{eq:emb_cap_5}
\end{align}

Equations \ref{eq:emb_cap_1}--\ref{eq:emb_cap_3} limit the capacity of EMBs allocated to the DRAM and SSD of SmartSSD $m$, while Equations \ref{eq:emb_cap_4} and \ref{eq:emb_cap_5} restrict the capacity of EMBs compressed in TT-format. In these constraints, the $tt\_cm_{j}$ used in Equation \ref{eq:emb_cap_4} refers to the compressed size, which depends on the percentage of the total EMB rows compressed in TT-format. Similar to the ICDF, this value is obtained through piecewise linear interpolation during DSA stage.

\textbf{Embedding Vector Access Latency}
To optimize processing overhead, constraints can be formulated to link the embedding vector access latency of each memory device to the batch size. This is achieved utilizing the PF obtained from the DSA and the variables derived from the three-level sharding process. The resulting equations are as follows:

\begin{equation}
\begin{aligned}
    (pf_{j} * BS) * (pct\_dra&m_{j} * t_{dram}) \\
    \hspace{4.1cm} & = c\_dram_{j} \quad \forall j \in J \label{eq:emb_latency_1}
\end{aligned}
\end{equation}

\vspace{-0.5cm}

\begin{equation}
    \hspace{0.8cm} (pf_{j} * BS) * (pct\_tt_{j} * t_{tt}) = c\_tt_{j} \quad \forall j \in J \label{eq:emb_latency_2}
\end{equation}

\vspace{-0.5cm}

\begin{equation}
\begin{aligned}
    \hspace{0.2cm} (pf_{j} * BS) * ((1 - pct\_dram&_{j} - pct\_tt_{j}) * t_{ssd}) \\
    & = c\_ssd_{j} \quad \forall j \in J \label{eq:emb_latency_3}
\end{aligned}
\end{equation}

\vspace{-0.5cm}

\begin{align}
    \hspace{1.2cm} \sum_{j} p_{mj} * c\_dram_{j} = c\_dram_{m} \quad \forall m &\in M \label{eq:emb_latency_4} \\
    \sum_{j} p_{mj} * c\_tt_{j} = c\_tt_{m} \quad \forall m &\in M \label{eq:emb_latency_5} \\
    \sum_{j} p_{mj} * c\_ssd_{j} = c\_ssd_{m} \quad \forall m &\in M \label{eq:emb_latency_6}
\end{align}

Equations \ref{eq:emb_latency_1}--\ref{eq:emb_latency_3} calculate latency cost of EMB $j$ allocated to DRAM, BRAM, and SSD. Additionally, Equations \ref{eq:emb_latency_4}--\ref{eq:emb_latency_6} define the constraints for determining the latency of each memory tier in the SmartSSD, using the $p_{mj}$ derived earlier in the EMB allocation process. By applying these constraints, a precise cost model for the embedding layer can be constructed, taking the specific characteristics of DLRM into account.

\textbf{MLP Layer Latency}
In addition to the embedding layer's cost constraints, the latency cost of the top MLP layer and the bottom MLP layer can also be formulated with the following constraints:

\begin{equation}
\begin{aligned}
    t_{mlp\_top} * (B&S / BS\_mini) \\
    & = c_{mlp\_top} * \sum_{m} (1-d_m) \quad \forall m \in M  \label{eq:mlp_cost_1}
\end{aligned}
\end{equation}

\vspace{-0.5cm}

\begin{equation}
\begin{aligned}
    t_{mlp\_bot} * (B&S / BS\_mini) \\
    & = c_{mlp\_bot} * \sum_{m} (1-d_m) \quad \forall m \in M \label{eq:mlp_cost_2}
\end{aligned}
\end{equation}

In the Equations \ref{eq:mlp_cost_1} and \ref{eq:mlp_cost_2}, $t_{mlp\_top}$ and $t_{mlp\_bot}$ are values obtained from the cycle-accurate MLP core simulator, and $BS\_mini$ refers to the batch size that can be processed by the MLP core in a single tile operation. The value of $BS\_mini$ is configurable, as the MLP core is equipped with multiple computing cores. The details of the hardware architecture will be discussed in Section \ref{subsec:core_arch}.

\textbf{Cost Definition}
To define the objective function for optimizing the DLRM, constraints for the previously mentioned variables were formulated. Using these constraints, the equations for the corresponding layers of the DLRM were established as follows:

\begin{align}
\hspace{0.4cm} \{ c\_dram_m, \, c\_tt_m, \, c\_ssd_m \} & \leq c_{emb} \quad \forall m \in M \label{eq:final_cost_1} \\
\{ c_{mlp\_bot}, \, c_{emb} \} & \leq c_{fnt} \label{eq:final_cost_2}
\end{align}

The constraint in Equation \ref{eq:final_cost_1} defines the cost of the embedding layer, $c_{emb}$, while the constraint in Equation \ref{eq:final_cost_2} defines $c_{fnt}$ which represents the maximum cost between the bottom MLP layer and the embedding layer. Using $c_{fnt}$ and $c_{mlp\_top}$, the objective function for Equation \ref{eq:obj_func} in the SRM can be formulated.

By constructing the cost model as described above, embedding layer partitioning and custom hardware core allocation can be achieved, enabling the implementation of an optimized hybrid-parallel processing system based on DLRM workload intensity.

\subsection{Address Remapping}
\label{subsec:soft_remap}

Address remapping is essential step for facilitating access to memory-allocated EMBs, which are partitioned by the SRM, using EMB access indices derived from input sparse features. The length of the remapping table generated in this stage matches the row length of the EMB. When the sparse feature index accesses the remapping table, a 32-bit remapped address is retrieved, which is then decoded to read the embedding vectors allocated in the corresponding memory devices. In detail, the remapped address is composed of \{device\_id[1:0], emb\_idx[29:0]\}, where a device\_id of 0 indicates the embedding vector is fetched from DRAM, 1 from BRAM, and 2 from SSD. By utilizing the remapping table loaded into a host, SCRec enables three-level sharding to be performed in a fine-grained manner.
\subsection{Hardware Accelerator Allocation}
\label{subsec:core_arch}

\begin{figure}[t]
    \centering
    \includegraphics[width=3.4in]{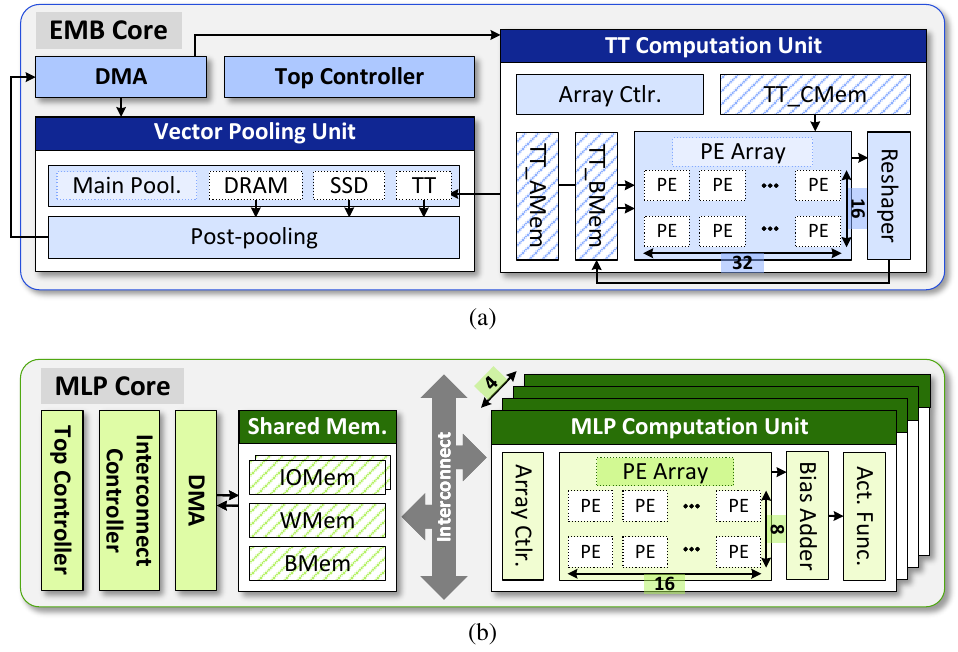}
    \caption{Hardware architecture of computing cores. (a) EMB core (b) MLP core}
    \label{fig:core_arch}
\end{figure}

The objective of the core mapper is to allocate either an MLP core for processing MLP layers or an EMB core for processing embedding layers based on $d_{m}$, derived from solving the cost model of SRM. The core mapper's applicability within SCRec is closely tied to the hardware architecture of the SmartSSD, which is equipped with an FPGA chip featuring reconfigurable hardware. This architecture enables adaptive programming of hardware configurations optimized for specific tasks prior to runtime. By leveraging this reconfigurable capability, hardware accelerators tailored to the DLRM workload can be mapped onto the SmartSSD.

\textbf{EMB Core}
EMB core fetches embedding vectors stored in TT-format, as well as those stored in DRAM and SSD, to perform pooling operations. It operates based on a 32-bit floating-point (FP) processing element (PE), as shown in Fig.\ref{fig:core_arch}(a). The key module of the EMB core is the TT computation unit (CU), which accelerates sequential matrix multiplication operations within the TT-core to generate approximated embedding vectors. While Equations \ref{eq:tt_rep} and \ref{eq:tt_emb} in Section \ref{sec:back_gnd} describe the calculation of a single element of an embedding vector, the TT CU computes the entire embedding vector simultaneously by processing the product of a 3-dimensional tensor using Equation \ref{eq:tt_emb_tensor}:

\begin{equation}
\begin{aligned}
    E(i,:) = \mathcal{G}_1(:, i_1,:,:) \mathcal{G}_2(:, i_2,:,:) \dots \mathcal{G}_d(:, i_d,:,:)
\label{eq:tt_emb_tensor}
\end{aligned}
\end{equation}

\begin{algorithm}[t]
\caption{TT-core Matrix-multiplication Sequence}
\label{alg:alg1}
\textbf{Input:} $G_k$ (TT-core of EMB table) \\
\textbf{Output:} $V$ (Approximated EMB vector) \\
\textbf{Parameter:} $K$ (TT-dim.), $R$ (TT-rank), $J$ (EMB dim.) \\
\begin{algorithmic} [1]

\STATE \textcolor{darkgreen}{/* Initialization */}
\FOR {$k = 1$ to $K$}
    \FOR {$i = 1$ to $I_k$}
        \IF {$k = 1$}
            \STATE $U_{k}[i] = Unfold(G_k[:,i,:,:], (-1, R))$
            \STATE $TT\_AMem = Fetch(U_{k}[i])$
        \ELSE
            \STATE $U_{k}[i] = Unfold(G_k[:,i,:,:], (R, -1))$
            \STATE $TT\_CMem = Fetch(U_{k}[i])$
        \ENDIF
    \ENDFOR
\ENDFOR
\STATE \textcolor{darkgreen}{/* Runtime computation */}
\FOR {$k = 1$ to $K$}
    \STATE \textcolor{darkgreen}{// Matrix multiplication}
    \IF {$k = 1$}
        \STATE $T_k = MatMul(U_k[i_k], U_{k+1}[i_{k+1}])$
    \ELSE
        \STATE $T_k = MatMul(R_k, U_{k+1}[i_{k+1}])$
    \ENDIF
    \STATE \textcolor{darkgreen}{// Matrix reshaping}
    \IF {$k \neq K$}
        \STATE $R_{k+1} = Reshape(T_k, (-1, rank))$
        \STATE $TT\_BMem = Fetch(R_{k+1})$
    \ELSE
        \STATE $V = Reshape(T_k, (J))$
    \ENDIF
\ENDFOR

\end{algorithmic}
\end{algorithm}

where TT-core \( \mathcal{G}_k \in \mathbb{R}^{R_{k-1} \times I_k \times J_k \times R_k} \) with \( R_0 = R_k = 1 \), the EMB row is defined as \( i = \sum_{k=1}^{d} i_k \prod_{l=k+1}^{d} I_l \) with the ranges \( 0 \leq i_d \leq I_d-1\). To perform 3D tensor multiplications on the EMB core, which is implemented with a 2D PE array, the 3D TT-core is unfolded into a 2D matrix and loaded into the on-chip memory during initialization. During sequential TT-core matrix multiplications, intermediate reshaping is performed to ensure conformability, as detailed in Algorithm \ref{alg:alg1}. Specifically, the multiply-accumulate (MAC) operations in the EMB core are executed using an output-stationary systolic array, processing the entire matrix in 16 $\times$ 32-sized output tiles iteratively. The input memory for the PE array is designed with dual channels in the column direction, effectively hiding the reshaping overhead within the memory read latency. Additionally, the vector pooling unit (VPU) independently performs pooling operations for each memory device, and the final pooled embedding vector is produced through the post-pooling module. The proposed EMB core not only accelerates TT computations but also supports pooling operations on embedding vectors stored across multi-tier memory devices, enabling complete embedding layer computations within the SmartSSD.

\begin{table}[t]
\centering
\caption{Experiment setup}
\label{tab:exp_setup}
\begin{tabular}{c|l}
\Xhline{2\arrayrulewidth}
\multicolumn{2}{c}{\textbf{Server}} \\ \Xhline{2\arrayrulewidth}
\textbf{CPU}              & Intel Xeon Silver 4310 @ 2.1GHz \\ \hline
\textbf{GPU}              & Nvidia A40 GDDR6 48GB \\ \hline
\textbf{DRAM}             & DDR4 SDRAM 256GB @ 3200Mbps \\ \Xhline{2\arrayrulewidth}
\multicolumn{2}{c}{\textbf{SSD Simulator (MQSim)}} \\ \Xhline{2\arrayrulewidth}
\textbf{Mem. Device}      & Samsung 860 EVO 3.84 TB \\ \hline
\textbf{NAND Flash}       & Samsung V-NAND V4 (TLC, 64-layer) \\ \hline
\textbf{PCIe Interface}   & Single Port PCIe Gen3 $\times$ 4 \\ \hline
\textbf{DRAM Cache}       & LPDDR4 SDRAM 2GB @ 1866Mbps \\ \hline
\textbf{Read Time ($t_R$)}& 45$\mu$s \\ \hline
\textbf{Page Size}        & 16KB \\ \Xhline{2\arrayrulewidth}
\multicolumn{2}{c}{\textbf{DRAM Simulator (Ramulator)}} \\ \Xhline{2\arrayrulewidth}
\textbf{Mem. Device}      & DDR4 SDRAM 4GB @ 2400Mbps \\ \Xhline{2\arrayrulewidth}
\multicolumn{2}{c}{\textbf{FPGA Chip}} \\ \Xhline{2\arrayrulewidth}
\textbf{Name}             & AMD Kintex™ Ultrascale+ KU15P \\ \hline
\textbf{On-chip Mem. Cap.}& BRAM - 4.325MB / URAM - 4.5MB \\ \hline
\textbf{\# of DSP Slices} & 1968 \\ \hline
\textbf{Core Freq.}       & 200MHz \\ \hline
\end{tabular}
\end{table}

\textbf{MLP Core}
The MLP core is designed to accelerate both the top and bottom MLP layers of the DLRM, utilizing a 32-bit FP PE, as illustrated in Fig.\ref{fig:core_arch}(b). The MLP CU consists of an 8 $\times$ 16 PE array, performing matrix multiplication operations in tiles using output-stationary computation. Input values for the MLP layer are read from IOMem and flow into the PE array in the column direction, while weight parameters are read from WMem and flow into the PE array in the row direction. Once partial outputs are computed in the PE array, the bias adder module adds bias values from BMem, and the activation function module applies a rectified linear unit (ReLU) operation before writing the results back to IOMem. The MLP core contains four CUs in total, with configurable input and weight data distribution from the interconnect, enabling the system to select between latency-optimized or throughput-optimized computation. Specifically, if input data is broadcast to all CUs and weight parameters are split so that each CU receives different weights, the MLP layer is processed in a latency-optimized manner. Conversely, if different input data is fed to each CU while weight parameters are broadcast to all CUs, the MLP layer is processed in a throughput-optimized manner. As a result, the MLP core in SCRec not only accelerates MLP operations but also supports configurable layer processing, significantly enhancing MLP processing performance.
\section{Evaluation}
\label{sec:eval}

\subsection{Experiment Setup}

We established an experimental environment using a GPU server and cycle-accurate simulators to evaluate the effectiveness of implementing DLRM in SCRec, with detailed specifications outlined in Table \ref{tab:exp_setup}. DLRM training and inference were conducted on a server equipped with Nvidia A40 GPUs to assess accuracy based on Meta DLRM\cite{github_dlrm} when applying the TT-format to the embedding layer. Additionally, we utilized Ramulator\cite{kim2015ramulator} and MQSim\cite{tavakkol2018mqsim}, simulators designed to evaluate the performance of DRAM and SSD, respectively, to measure processing performance during SCRec's runtime operation. To compare SCRec with a multi-GPU system, we measured the power consumption of SSD and DRAM using SimpleSSD\cite{jung2017simplessd} and VAMPIRE\cite{ghose2018your}. FPGA power consumption was profiled using the Xilinx Vivado power analysis tool to evaluate the power of the implemented core logic. We also designed an in-house cycle-accurate core simulator, similar to the memory simulators, to measure performance for simulating the hardware design implemented on the FPGA chip. To synchronize the memory simulators and the core simulator, modifications were made to enable transmission and reception of control signals and data between simulators via inter-process communication (IPC) using the TCP protocol\cite{cerf1974protocol}. This synchronization ensured accurate emulation of real-time system behavior by aligning the simulators at each cycle.

\begin{table}[t]
\centering
\caption{DLRM evaluation dataset specification}
\label{tab:data_spec}
\begin{tabular}{c|c|c|c}
    \Xhline{2\arrayrulewidth}
    \multirow{2}{*}{\textbf{Dataset}} & \multirow{2}{*}{Criteo Kaggle} & \multicolumn{2}{c}{Meta Synthetic} \\ \cline{3-4}
                             &                                & 2021 & 2022 \\ \Xhline{2\arrayrulewidth}
    \textbf{MLP Layer}                & O           & X          & X    \\ \hline
    \textbf{\# of EMB Table}          & 26          & 856        & 788  \\ \hline
    \textbf{Avg. EMB Rows}            & 1,298,560   & 2,720,716    & 4,841,017 \\ \hline
    \textbf{Avg. PF}                  & 1           & 8.34       & 13.6  \\ \hline
    \textbf{\# of Data Sample}        & 45,840,617  & 65,536     & 131,072   \\ \Xhline{2\arrayrulewidth}
\end{tabular}
\end{table}

Regarding the datasets used for evaluation, large-scale industrial DLRM datasets are not publicly available. Therefore, we utilized the Criteo Display Advertising (CDA) dataset\cite{dataset_criteo_kaggle} and the MELS dataset\cite{github_meta_syn}. As the characteristics of these datasets differ, as shown in Table \ref{tab:data_spec}, the appropriate dataset was selected based on the evaluation metrics. Specifically, the CDA dataset supports the evaluation of the entire DLRM, including both the embedding and MLP layers. However, due to significant differences in the size and access patterns of its embedding layer compared to industrial DLRMs, it was primarily used for assessing DLRM accuracy and the performance of acceleration cores in our evaluation. Conversely, while the MELS dataset is limited to evaluating the embedding layer without the MLP layers, it offers insights into addressing key challenges currently faced by industrial DLRMs. Consequently, we focused on using the MELS dataset to evaluate the energy efficiency and performance of the industrial DLRM implementation, with a specific emphasis on the embedding layer.

To evaluate the SCRec software framework, we derived statistical parameters for the SRM from the DSA using subsampled data from the CDA and MELS datasets. For the CDA dataset, a 1\% randomly subsampled portion was used to extract these parameters. In contrast, for the MELS dataset, given its smaller data sample size, we increased the subsampling rate to 10\% to enhance statistical reliability. In the SRM, the Gurobi optimizer\cite{gurobi_optimizer} was employed to construct the MIP-based cost model. Specifically, the $hot\_thr$ was empirically set to 1 if its value was below 0.01\% of the largest EMB row in the entire EMB and to 0.99 if it exceeded this threshold. To generate TT-formatted EMBs, the t3nsor library\cite{hrinchuk2019tensorized} was utilized, applying a TT-rank of 4 to achieve higher data compression performance.

\subsection{Hardware Implementation Results}

\begin{figure}[t]
    \centering
    \includegraphics[width=3.4in]{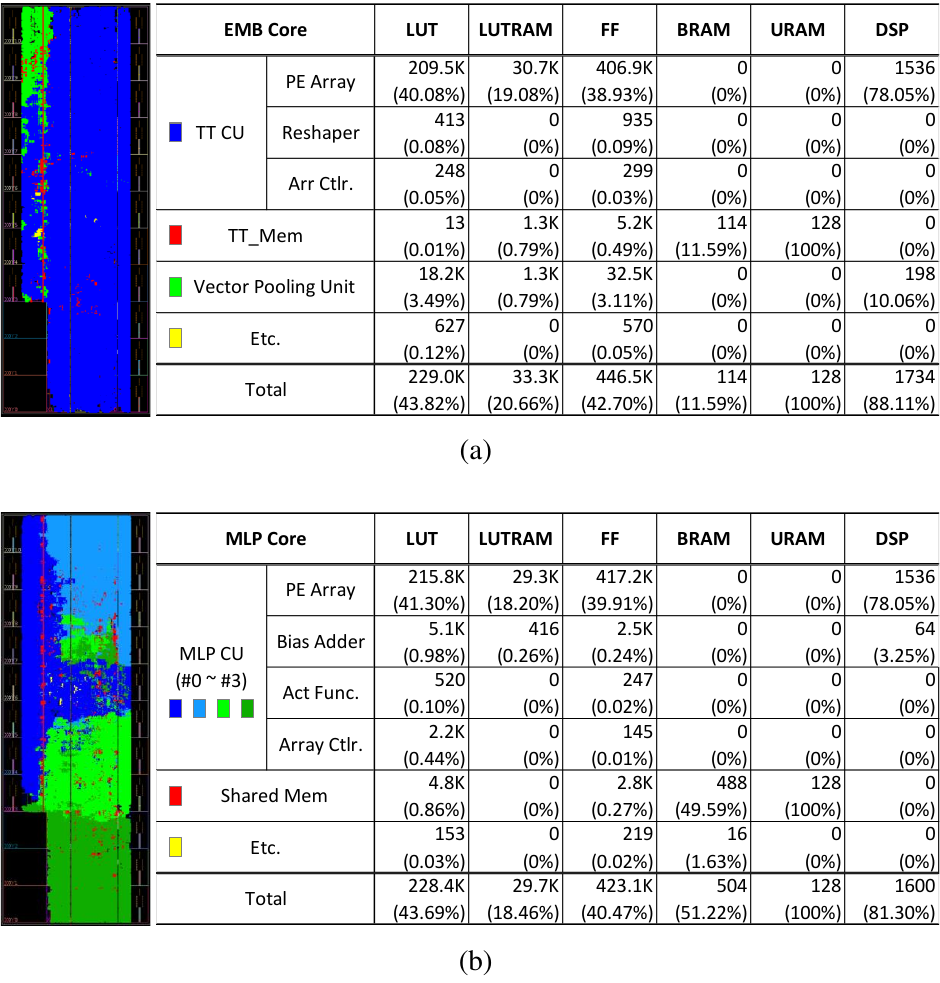}
    \caption{FPGA layout and resource utilization on SmartSSD. (a) EMB core (b) MLP core}
    \label{fig:fpga_util}
\end{figure}

To verify the feasibility of implementing the hardware cores of SCRec on a SmartSSD, we synthesized and performed place-and-route operations for the computation kernels of the EMB core and MLP core at a target frequency of 200 MHz. The implementation details and results are shown in Fig.\ref{fig:fpga_util}(a) and Fig.\ref{fig:fpga_util}(b), respectively. Both cores demonstrated the highest utilization in the CU module, primarily due to the hardware computation data format being based on single-precision FP. To achieve this, the PE modules instantiate FP multipliers and FP accumulators synthesized using Xilinx’s FP Intellectual Property (IP). This design optimizes hardware by utilizing approximately 78\% of the DSP slices, leveraging look-up tables (LUTs) and flip-flops (FFs). For the EMB core, the vector pooling unit (VPU) pools embedding vectors read from each memory device using FP adders, resulting in about 10\% DSP slice utilization. Additionally, the post-pooling unit within the VPU computes the average pooling vector using an FP divider. In the MLP core, the bias adder module performs vector addition between the matrix multiplication result and the bias parameter using the FP adder module, which utilizes approximately 3\% of the DSP slices. The difference in DSP slice utilization between non-CU modules and the CU modules arises from their differing workloads; non-CU modules perform element-wise operations, resulting in a lower computational load. As a result, these modules are designed to consume relatively less hardware resources.

\subsection{Performance}
\begin{figure}[t]
    \centering
    \includegraphics[width=3.4in]{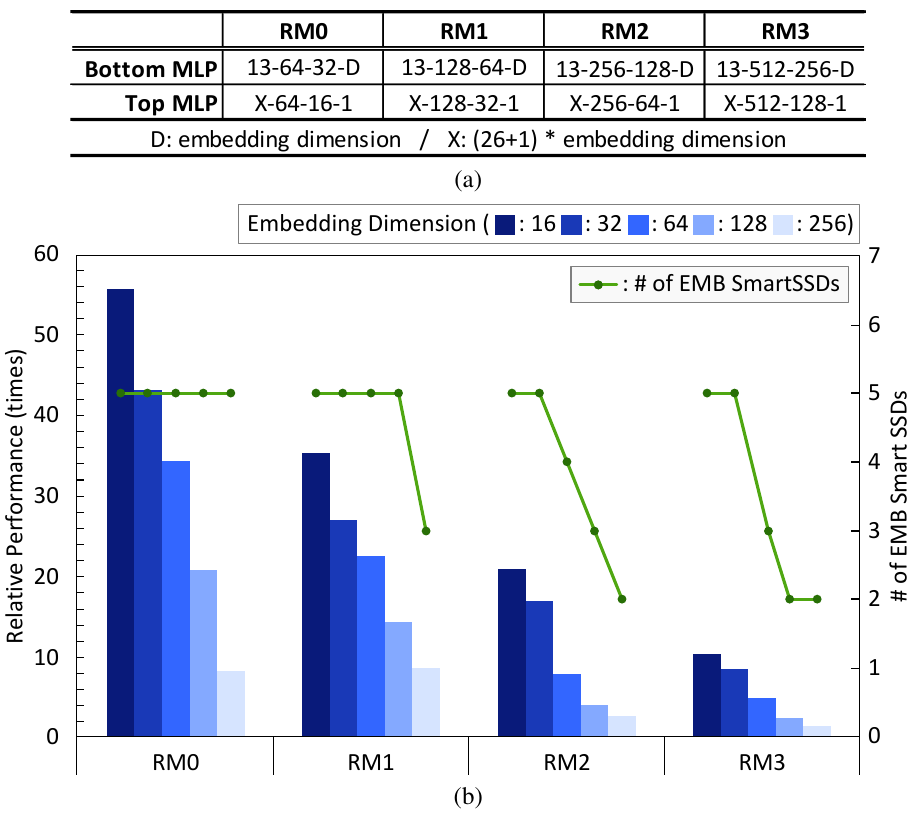}
    \caption{Performance comparison between CPU-DRAM system and SCRec. (a) DLRM configuration (b) Relative performance and EMB core allocation}
    \label{fig:eval_perf}
\end{figure}

We evaluated SCRec using the CDA dataset by analyzing core allocation variations based on the layer workload in the DLRM and comparing its performance to a CPU-DRAM system. This evaluation utilized the inference per second (IPS) metric with a batch size of 128, using 8 SmartSSDs. The DLRM configuration applied four different setups, as described in Fig.\ref{fig:eval_perf}(a), with each model containing an embedding layer composed of 26 EMBs. In the DLRM, the input layer for the top MLP layer was determined by concatenating the outputs of the 26 EMBs and the bottom MLP output, following the constraints of the Meta DLRM \cite{github_dlrm}. Analyzing the graph in Fig.\ref{fig:eval_perf}(b), SCRec significantly outperforms the CPU-DRAM system when the embedding layer workload exceeds that of the MLP layers. Specifically, for an embedding dimension of 16, SCRec shows a performance improvement of 10.37$\times$-55.77$\times$ across the four recommendation models (RMs). This enhancement is attributed to SCRec’s ability to leverage multiple SmartSSD devices, enabling model parallelism in processing the embedding layer and thereby increasing processing bandwidth compared to the CPU-DRAM system. Within each RM, as the embedding dimension increases, the relative performance of SCRec declines. This is because a larger embedding dimension increases the processing overhead of the top MLP layer, indicating that SCRec is more effective for embedding-dominant DLRM workloads than for MLP-dominant ones. Regarding core allocation for the SmartSSDs, as the workload of the MLP layer grows relative to that of the embedding layer, the number of SmartSSDs assigned to EMB cores decreases, while the number assigned to MLP cores increases. This demonstrates that by optimizing the cost model in its software stack, SCRec effectively mitigates low core utilization caused by imbalances in processing overheads between the embedding and MLP layers. These results indicate that SCRec can optimize the system based on workload intensity across various DLRM configurations and improve processing performance through hybrid-parallel processing.

\subsection{Energy Efficiency}

\begin{figure}[t]
    \centering
    \includegraphics[width=3.4in]{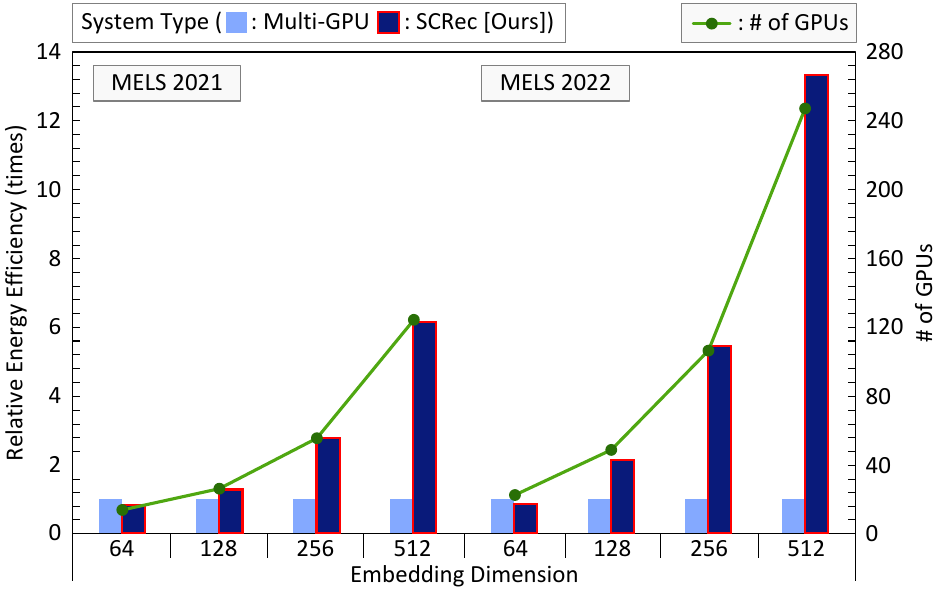}
    \caption{Energy efficiency comparison between multi-GPU system and SCRec.}
    \label{fig:eval_energy}
\end{figure}

We evaluated the energy efficiency of SCRec, deployed on 8 SmartSSDs, against a multi-GPU system by varying embedding dimensions and utilizing the MELS 2021 and 2022 datasets. The batch size was set to 1024, and model parallelism was applied similarly in both systems, with EMBs split table-wise and distributed to GPUs in a round-robin manner based on their index. Due to the large size of the model parameters in MELS, which could not fit into a single server, a hash table was applied to the EMBs during measurement to limit the maximum index of the EMBs. Energy efficiency was measured using the IPS per watt (IPS/W) metric to assess relative performance. As shown in Fig.\ref{fig:eval_energy}, SCRec demonstrated superior energy efficiency compared to the multi-GPU system across both datasets, except when the embedding dimension was set to 64. As the embedding dimension increased, the improvement in energy efficiency became more pronounced. Specifically, with an embedding dimension of 512, SCRec achieved energy efficiency improvements of 6.14$\times$ for the 2021 dataset and 13.35$\times$ for the 2022 dataset. This improvement stems from the fact that as the embedding dimension increases, the number of GPUs required to store DLRM parameters grows significantly. In contrast, SCRec can fully accommodate DLRM parameters within the 32 TB capacity of the SmartSSDs. Moreover, SCRec compensates for hot embedding vectors that cannot be fetched to DRAM due to capacity limitations by compressing them using the TT-format, enabling GB/s-level bandwidth. These findings highlight that SCRec provides an energy-efficient solution for the embedding layers of industrial-scale DLRMs compared to multi-GPU systems. Unlike systems requiring dozens or even hundreds of GPUs with complex network topologies, SCRec achieves energy efficiency by utilizing multiple SmartSSD devices within a single server, eliminating network communication overhead.

\subsection{Ablation Study}

\begin{figure}[t]
    \centering
    \includegraphics[width=3.4in]{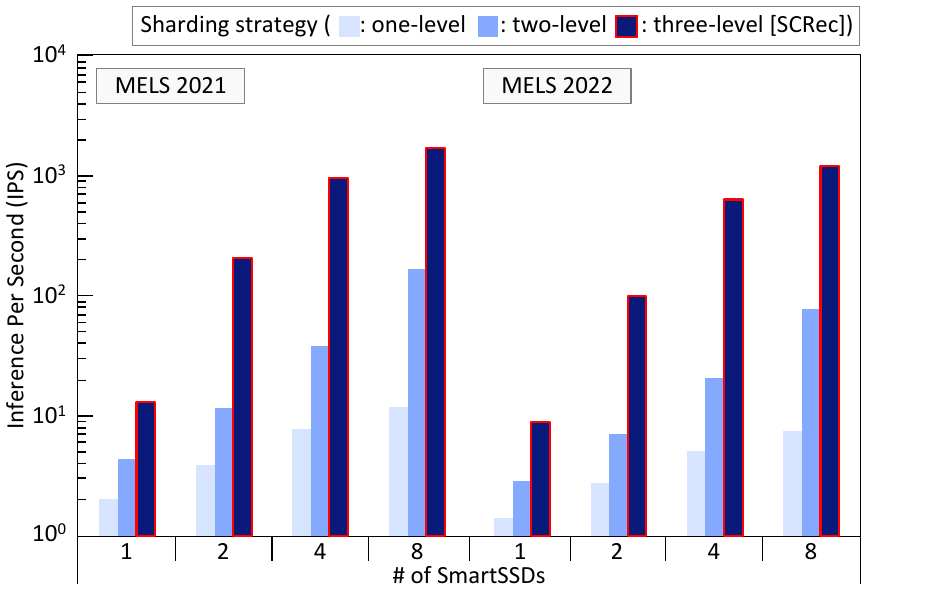}
    \caption{Performance-based ablation study using Meta embedding lookup synthetic dataset.}
    \label{fig:eval_abl}
\end{figure}

We evaluated the performance changes in SCRec based on the sharding levels applied, varying the number of SmartSSDs and using the MELS 2021 and 2022 datasets, which include only embedding layer access patterns. An embedding dimension of 256 and a batch size of 1024 were used. For one-level sharding, only SSD was utilized, while for two-level sharding, both SSD and DRAM were employed. As shown in Fig.\ref{fig:eval_abl}, increasing the sharding level by incorporating additional memory devices to store EMBs, while maintaining the same number of SmartSSDs, leads to improved inference performance. For the 2021 dataset, the three-level sharding scheme applied in SCRec achieved performance improvements of 141.34$\times$ and 10.11$\times$ over one-level and two-level sharding, respectively, when 8 SmartSSDs were used. With a single SmartSSD, the improvements were 6.55$\times$ and 3.02$\times$, respectively. Similarly, for the 2022 dataset, three-level sharding delivered performance gains of 161.77$\times$ and 15.48$\times$ over one-level and two-level sharding, respectively, with 8 SmartSSDs, and 6.31$\times$ and 3.14$\times$ with a single SmartSSD. The variation in performance improvements with higher sharding levels, while keeping the number of SmartSSDs constant, can be attributed to several factors. Transitioning from one-level to two-level sharding allows frequently accessed embedding vectors to be fetched from DRAM, mitigating the performance degradation caused by the lower bandwidth of SSDs. Advancing from two-level to three-level sharding compensates for DRAM’s capacity limitations by storing highly compressed embedding vectors in TT-format. This approach effectively hides SSD access latency by overlapping it with the access latencies of DRAM and BRAM, resulting in superior performance with three-level sharding. Moreover, increasing the number of SmartSSDs in SCRec’s three-level sharding further enhances performance due to the application of model parallelism in the embedding layer. As the number of devices grows, the effective bandwidth increases proportionally. Notably, for both datasets, the most significant performance improvements occur when the number of SmartSSDs increases from 1 to 2, with gains of 17.64$\times$ and 11.46$\times$, respectively. This is because fewer SmartSSDs exacerbate the capacity limitations when fetching hot embedding vectors into high-bandwidth memory devices. These results demonstrate that three-level sharding in SCRec not only addresses DRAM capacity limitations but also effectively hides SSD access latency, significantly enhancing the performance of the embedding layer in industrial-scale DLRMs.

\subsection{Accuracy}

\begin{figure}[t]
    \centering
    \includegraphics[width=3.4in]{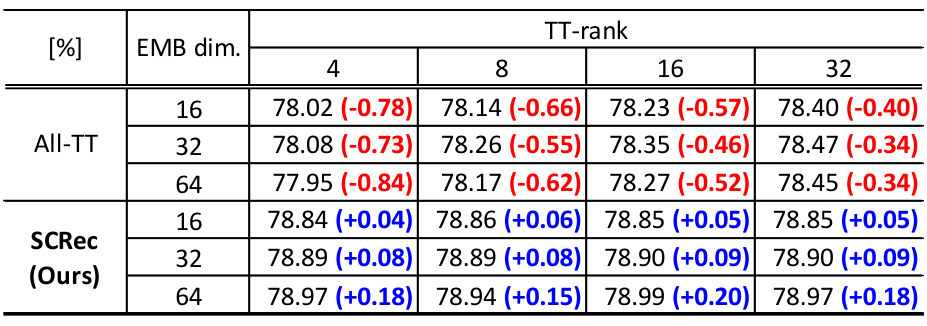}
    \caption{Accuracy comparison. (Numbers in parentheses represent accuracy difference from vanilla-DLRM)}
    \label{fig:eval_accu}
\end{figure}

Using low-rank approximation techniques such as TT decomposition significantly increases the data CR. However, applying the TT-format introduces information loss, and as the TT-rank decreases, the quality of data representation deteriorates, resulting in lower model accuracy. Therefore, minimizing accuracy drop becomes a critical focus when compressing the embedding layer in the DLRM using the TT-format. To evaluate accuracy when part of the EMBs is represented in the TT-format with SCRec, we varied the embedding dimension and rank, as shown in Fig.\ref{fig:eval_accu}. For this evaluation, we used the RM3 configuration described in Fig.\ref{fig:eval_perf}, based on the CDA dataset. When all embedding layers were replaced with the TT-format, and inference accuracy was measured after training, the accuracy drop was highest at a TT-rank of 4 across all embedding dimensions, ranging from -0.84\% to -0.73\%. In contrast, the accuracy drop was lowest at a TT-rank of 32, ranging from -0.40\% to -0.34\%, demonstrating that higher TT-ranks improve the data quality of EMBs and reduce accuracy loss. SCRec, on the other hand, exhibited a slight accuracy increase of +0.04\% to +0.20\% across all experiments. This improvement occurs because SCRec does not apply the TT-format to all EMBs. Instead, it stores most frequently accessed embedding vectors in DRAM, while less frequently accessed ones are stored in the TT-format. Consequently, only a small portion of embedding vector accesses are impacted by accuracy loss. These findings indicate that in SCRec, using a TT-rank of 4 achieves high compression rates while maintaining high-throughput access to the embedding layer without any accuracy drop. Beyond the CDA dataset, SCRec facilitates partial compression of EMBs in large-scale industrial DLRMs using the TT-format, suggesting its potential for effective application in these scenarios while mitigating accuracy loss.

\section{Related Works}
\label{sec:rel_work}

\subsection {Multi-tier Memory Caching and Sharding}

Due to the massive size of the embedding layer and the irregular access patterns with a power-law distribution in DLRM, many studies have focused on leveraging multi-tier memory and caching schemes to reduce EMB access latency. Scratch-Pipe\cite{kwon2022training} utilizes a 2-level memory hierarchy consisting of host CPU memory and GPU memory, caching embedding vectors in GPU memory to overcome the limited memory capacity of GPUs. This approach enables the embedding layer to be trained at GPU memory speed. AIBox\cite{zhao2019aibox} integrates GPU and SSD devices to store large embedding layers, implementing an SSD cache management system to provide low-latency access to embedding layers stored in SSDs. This allows industrial-scale recommendation models to be trained on a single node.

In addition to caching methods, sharding schemes that split model parameters across multiple devices have also been extensively researched. RecShard\cite{sethi2022recshard} leverages multiple GPUs, placing frequently accessed data in the GPU’s HBM and less frequently accessed data in unified virtual memory (UVM), which encompasses host DRAM and GPU HBM. By employing a cost model before runtime, RecShard effectively partitions the DLRM across GPU devices, significantly enhancing training throughput. Similarly, AutoShard\cite{zha2022autoshard} utilizes a cost model to estimate table computation costs and incorporates deep reinforcement learning (DRL) to address the EMB partitioning challenge across multiple GPUs, achieving balanced sharding among GPUs.

\subsection {Tensor-train Decomposition}

In addition to the challenges posed by large EMB sizes in DLRM, the substantial vocabulary size in transformer models presents a significant hurdle in natural language processing (NLP). To address this, TT-embedding\cite{hrinchuk2019tensorized} was introduced to efficiently compress EMBs using tensor-train (TT) decomposition\cite{oseledets2011tensor}. This method enables seamless integration into any deep learning framework and supports training via backpropagation. TT-Rec\cite{yin2021tt} applies the TT-embedding format to effectively compress EMBs in DLRM, introducing a scheme that initializes TT-core weights based on a sampled Gaussian distribution. It demonstrated negligible accuracy loss when tested on the CDA and Criteo Terabyte datasets. Similarly, EL-Rec\cite{wang2022rec} utilized the TT-format to compress the embedding layer and developed a TT-based training pipeline system. This approach mitigated host communication latency between the CPU and GPU, improving training performance on a single GPU.

\section{Conclusion}
\label{sec:conc}

In this paper, we present SCRec, a scalable computational storage system with statistical sharding and TT decomposition for recommendation models. On the software side, SCRec leverages three-level statistical sharding to meet the high bandwidth requirements of DLRM by storing hot data in high-bandwidth memory devices and cold data in a low-bandwidth memory device. Additionally, SCRec can adaptively configure memory-centric and compute-centric cores based on DLRM workload intensity, thereby configuring optimized system. On the hardware side, SCRec implements custom hardware acceleration cores to enhance DLRM computations. In particular, it enables the high-performance reconstruction of approximated embedding vectors from significantly compressed TT-format, complementing memory capacity limitations while enhancing memory bandwidth. By integrating the software framework with the hardware accelerators into our system using multiple SmartSSDs, SCRec achieved up to 55.77$\times$ inference performance improvement compared to a CPU-DRAM system and up to 13.35$\times$ energy efficiency improvement compared to a multi-GPU system. These results validate that large-scale DLRMs can be implemented on a single server with high performance and low energy costs, eliminating data communication overhead.

\bibliographystyle{IEEEtran}
\bibliography{IEEEabrv, 99_references}

\begin{thebibliography}{10}
\providecommand{\url}[1]{#1}
\csname url@samestyle\endcsname
\providecommand{\newblock}{\relax}
\providecommand{\bibinfo}[2]{#2}
\providecommand{\BIBentrySTDinterwordspacing}{\spaceskip=0pt\relax}
\providecommand{\BIBentryALTinterwordstretchfactor}{4}
\providecommand{\BIBentryALTinterwordspacing}{\spaceskip=\fontdimen2\font plus
\BIBentryALTinterwordstretchfactor\fontdimen3\font minus \fontdimen4\font\relax}
\providecommand{\BIBforeignlanguage}[2]{{%
\expandafter\ifx\csname l@#1\endcsname\relax
\typeout{** WARNING: IEEEtran.bst: No hyphenation pattern has been}%
\typeout{** loaded for the language `#1'. Using the pattern for}%
\typeout{** the default language instead.}%
\else
\language=\csname l@#1\endcsname
\fi
#2}}
\providecommand{\BIBdecl}{\relax}
\BIBdecl

\bibitem{resnick1997recommender}
P.~Resnick and H.~R. Varian, ``Recommender systems,'' \emph{Communications of the ACM}, vol.~40, no.~3, pp. 56--58, 1997.

\bibitem{raghuwanshi2019recommendation}
S.~K. Raghuwanshi and R.~K. Pateriya, ``Recommendation systems: techniques, challenges, application, and evaluation,'' in \emph{Soft Computing for Problem Solving: SocProS 2017, Volume 2}.\hskip 1em plus 0.5em minus 0.4em\relax Springer, 2019, pp. 151--164.

\bibitem{darban2022ghrs}
Z.~Z. Darban and M.~H. Valipour, ``Ghrs: Graph-based hybrid recommendation system with application to movie recommendation,'' \emph{Expert Systems with Applications}, vol. 200, p. 116850, 2022.

\bibitem{covington2016deep}
P.~Covington, J.~Adams, and E.~Sargin, ``Deep neural networks for youtube recommendations,'' in \emph{Proceedings of the 10th ACM conference on recommender systems}, 2016, pp. 191--198.

\bibitem{cheng2016wide}
H.-T. Cheng, L.~Koc, J.~Harmsen, T.~Shaked, T.~Chandra, H.~Aradhye, G.~Anderson, G.~Corrado, W.~Chai, M.~Ispir \emph{et~al.}, ``Wide \& deep learning for recommender systems,'' in \emph{Proceedings of the 1st workshop on deep learning for recommender systems}, 2016, pp. 7--10.

\bibitem{naumov2020deep}
M.~Naumov, J.~Kim, D.~Mudigere, S.~Sridharan, X.~Wang, W.~Zhao, S.~Yilmaz, C.~Kim, H.~Yuen, M.~Ozdal \emph{et~al.}, ``Deep learning training in facebook data centers: Design of scale-up and scale-out systems,'' \emph{arXiv preprint arXiv:2003.09518}, 2020.

\bibitem{goldberg1992using}
D.~Goldberg, D.~Nichols, B.~M. Oki, and D.~Terry, ``Using collaborative filtering to weave an information tapestry,'' \emph{Communications of the ACM}, vol.~35, no.~12, pp. 61--70, 1992.

\bibitem{hu2008collaborative}
Y.~Hu, Y.~Koren, and C.~Volinsky, ``Collaborative filtering for implicit feedback datasets,'' in \emph{2008 Eighth IEEE international conference on data mining}.\hskip 1em plus 0.5em minus 0.4em\relax Ieee, 2008, pp. 263--272.

\bibitem{koren2009matrix}
Y.~Koren, R.~Bell, and C.~Volinsky, ``Matrix factorization techniques for recommender systems,'' \emph{Computer}, vol.~42, no.~8, pp. 30--37, 2009.

\bibitem{sarwar2001item}
B.~Sarwar, G.~Karypis, J.~Konstan, and J.~Riedl, ``Item-based collaborative filtering recommendation algorithms,'' in \emph{Proceedings of the 10th international conference on World Wide Web}, 2001, pp. 285--295.

\bibitem{naumov2019deep}
M.~Naumov, D.~Mudigere, H.-J.~M. Shi, J.~Huang, N.~Sundaraman, J.~Park, X.~Wang, U.~Gupta, C.-J. Wu, A.~G. Azzolini \emph{et~al.}, ``Deep learning recommendation model for personalization and recommendation systems,'' \emph{arXiv preprint arXiv:1906.00091}, 2019.

\bibitem{shi2020compositional}
H.-J.~M. Shi, D.~Mudigere, M.~Naumov, and J.~Yang, ``Compositional embeddings using complementary partitions for memory-efficient recommendation systems,'' in \emph{Proceedings of the 26th ACM SIGKDD International Conference on Knowledge Discovery \& Data Mining}, 2020, pp. 165--175.

\bibitem{gupta2020architectural}
U.~Gupta, C.-J. Wu, X.~Wang, M.~Naumov, B.~Reagen, D.~Brooks, B.~Cottel, K.~Hazelwood, M.~Hempstead, B.~Jia \emph{et~al.}, ``The architectural implications of facebook's dnn-based personalized recommendation,'' in \emph{2020 IEEE International Symposium on High Performance Computer Architecture (HPCA)}.\hskip 1em plus 0.5em minus 0.4em\relax IEEE, 2020, pp. 488--501.

\bibitem{firoozshahian2023mtia}
A.~Firoozshahian, J.~Coburn, R.~Levenstein, R.~Nattoji, A.~Kamath, O.~Wu, G.~Grewal, H.~Aepala, B.~Jakka, B.~Dreyer \emph{et~al.}, ``Mtia: First generation silicon targeting meta's recommendation systems,'' in \emph{Proceedings of the 50th Annual International Symposium on Computer Architecture}, 2023, pp. 1--13.

\bibitem{agarwal2023bagpipe}
S.~Agarwal, C.~Yan, Z.~Zhang, and S.~Venkataraman, ``Bagpipe: Accelerating deep recommendation model training,'' in \emph{Proceedings of the 29th Symposium on Operating Systems Principles}, 2023, pp. 348--363.

\bibitem{kwon2019tensordimm}
Y.~Kwon, Y.~Lee, and M.~Rhu, ``Tensordimm: A practical near-memory processing architecture for embeddings and tensor operations in deep learning,'' in \emph{Proceedings of the 52nd Annual IEEE/ACM International Symposium on Microarchitecture}, 2019, pp. 740--753.

\bibitem{gupta2021training}
V.~Gupta, D.~Choudhary, P.~Tang, X.~Wei, X.~Wang, Y.~Huang, A.~Kejariwal, K.~Ramchandran, and M.~W. Mahoney, ``Training recommender systems at scale: Communication-efficient model and data parallelism,'' in \emph{Proceedings of the 27th ACM SIGKDD Conference on Knowledge Discovery \& Data Mining}, 2021, pp. 2928--2936.

\bibitem{zhao2020distributed}
W.~Zhao, D.~Xie, R.~Jia, Y.~Qian, R.~Ding, M.~Sun, and P.~Li, ``Distributed hierarchical gpu parameter server for massive scale deep learning ads systems,'' \emph{Proceedings of Machine Learning and Systems}, vol.~2, pp. 412--428, 2020.

\bibitem{sethi2022recshard}
G.~Sethi, B.~Acun, N.~Agarwal, C.~Kozyrakis, C.~Trippel, and C.-J. Wu, ``Recshard: statistical feature-based memory optimization for industry-scale neural recommendation,'' in \emph{Proceedings of the 27th ACM International Conference on Architectural Support for Programming Languages and Operating Systems}, 2022, pp. 344--358.

\bibitem{zhao2019aibox}
W.~Zhao, J.~Zhang, D.~Xie, Y.~Qian, R.~Jia, and P.~Li, ``Aibox: Ctr prediction model training on a single node,'' in \emph{Proceedings of the 28th ACM International Conference on Information and Knowledge Management}, 2019, pp. 319--328.

\bibitem{gupta2020deeprecsys}
U.~Gupta, S.~Hsia, V.~Saraph, X.~Wang, B.~Reagen, G.-Y. Wei, H.-H.~S. Lee, D.~Brooks, and C.-J. Wu, ``Deeprecsys: A system for optimizing end-to-end at-scale neural recommendation inference,'' in \emph{2020 ACM/IEEE 47th Annual International Symposium on Computer Architecture (ISCA)}.\hskip 1em plus 0.5em minus 0.4em\relax IEEE, 2020, pp. 982--995.

\bibitem{sun2022rm}
X.~Sun, H.~Wan, Q.~Li, C.-L. Yang, T.-W. Kuo, and C.~J. Xue, ``Rm-ssd: In-storage computing for large-scale recommendation inference,'' in \emph{2022 IEEE International Symposium on High-Performance Computer Architecture (HPCA)}.\hskip 1em plus 0.5em minus 0.4em\relax IEEE, 2022, pp. 1056--1070.

\bibitem{kim2020reducing}
M.~Kim and S.~Lee, ``Reducing tail latency of dnn-based recommender systems using in-storage processing,'' in \emph{Proceedings of the 11th ACM SIGOPS Asia-Pacific Workshop on Systems}, 2020, pp. 90--97.

\bibitem{wilkening2021recssd}
M.~Wilkening, U.~Gupta, S.~Hsia, C.~Trippel, C.-J. Wu, D.~Brooks, and G.-Y. Wei, ``Recssd: near data processing for solid state drive based recommendation inference,'' in \emph{Proceedings of the 26th ACM International Conference on Architectural Support for Programming Languages and Operating Systems}, 2021, pp. 717--729.

\bibitem{wan2021flashembedding}
H.~Wan, X.~Sun, Y.~Cui, C.-L. Yang, T.-W. Kuo, and C.~J. Xue, ``Flashembedding: storing embedding tables in ssd for large-scale recommender systems,'' in \emph{Proceedings of the 12th ACM SIGOPS Asia-Pacific Workshop on Systems}, 2021, pp. 9--16.

\bibitem{wang2022merlin}
Z.~Wang, Y.~Wei, M.~Lee, M.~Langer, F.~Yu, J.~Liu, S.~Liu, D.~G. Abel, X.~Guo, J.~Dong \emph{et~al.}, ``Merlin hugectr: Gpu-accelerated recommender system training and inference,'' in \emph{Proceedings of the 16th ACM Conference on Recommender Systems}, 2022, pp. 534--537.

\bibitem{xiao2023g}
Y.~Xiao, S.~Zhao, Z.~Zhou, Z.~Huan, L.~Ju, X.~Zhang, L.~Wang, and J.~Zhou, ``G-meta: Distributed meta learning in gpu clusters for large-scale recommender systems,'' in \emph{Proceedings of the 32nd ACM International Conference on Information and Knowledge Management}, 2023, pp. 4365--4369.

\bibitem{mudigere2022software}
D.~Mudigere, Y.~Hao, J.~Huang, Z.~Jia, A.~Tulloch, S.~Sridharan, X.~Liu, M.~Ozdal, J.~Nie, J.~Park \emph{et~al.}, ``Software-hardware co-design for fast and scalable training of deep learning recommendation models,'' in \emph{Proceedings of the 49th Annual International Symposium on Computer Architecture}, 2022, pp. 993--1011.

\bibitem{wang2024rap}
Z.~Wang, Y.~Wang, J.~Deng, D.~Zheng, A.~Li, and Y.~Ding, ``Rap: Resource-aware automated gpu sharing for multi-gpu recommendation model training and input preprocessing,'' in \emph{Proceedings of the 29th ACM International Conference on Architectural Support for Programming Languages and Operating Systems, Volume 2}, 2024, pp. 964--979.

\bibitem{kwon2021tensor}
Y.~Kwon, Y.~Lee, and M.~Rhu, ``Tensor casting: Co-designing algorithm-architecture for personalized recommendation training,'' in \emph{2021 IEEE International Symposium on High-Performance Computer Architecture (HPCA)}.\hskip 1em plus 0.5em minus 0.4em\relax IEEE, 2021, pp. 235--248.

\bibitem{oseledets2011tensor}
I.~V. Oseledets, ``Tensor-train decomposition,'' \emph{SIAM Journal on Scientific Computing}, vol.~33, no.~5, pp. 2295--2317, 2011.

\bibitem{yin2021tt}
C.~Yin, B.~Acun, C.-J. Wu, and X.~Liu, ``Tt-rec: Tensor train compression for deep learning recommendation models,'' \emph{Proceedings of Machine Learning and Systems}, vol.~3, pp. 448--462, 2021.

\bibitem{github_meta_syn}
\BIBentryALTinterwordspacing
{{Facebook Research}}, ``Embedding lookup synthetic dataset,'' 2021, accessed: 2021-12-09. [Online]. Available: \url{https://github.com/facebookresearch/dlrm_datasets}
\BIBentrySTDinterwordspacing

\bibitem{github_dlrm}
\BIBentryALTinterwordspacing
{Facebook Research}, ``An implementation of a deep learning recommendation model (dlrm),'' 2019, accessed: 2019-03-21. [Online]. Available: \url{https://github.com/facebookresearch/dlrm}
\BIBentrySTDinterwordspacing

\bibitem{kim2015ramulator}
Y.~Kim, W.~Yang, and O.~Mutlu, ``Ramulator: A fast and extensible dram simulator,'' \emph{IEEE Computer architecture letters}, vol.~15, no.~1, pp. 45--49, 2015.

\bibitem{tavakkol2018mqsim}
A.~Tavakkol, J.~G{\'o}mez-Luna, M.~Sadrosadati, S.~Ghose, and O.~Mutlu, ``$\{$MQSim$\}$: A framework for enabling realistic studies of modern $\{$Multi-Queue$\}$$\{$SSD$\}$ devices,'' in \emph{16th USENIX Conference on File and Storage Technologies (FAST 18)}, 2018, pp. 49--66.

\bibitem{jung2017simplessd}
M.~Jung, J.~Zhang, A.~Abulila, M.~Kwon, N.~Shahidi, J.~Shalf, N.~S. Kim, and M.~Kandemir, ``Simplessd: Modeling solid state drives for holistic system simulation,'' \emph{IEEE Computer Architecture Letters}, vol.~17, no.~1, pp. 37--41, 2017.

\bibitem{ghose2018your}
S.~Ghose, A.~G. Yaglik{\c{c}}i, R.~Gupta, D.~Lee, K.~Kudrolli, W.~X. Liu, H.~Hassan, K.~K. Chang, N.~Chatterjee, A.~Agrawal \emph{et~al.}, ``What your dram power models are not telling you: Lessons from a detailed experimental study,'' \emph{Proceedings of the ACM on Measurement and Analysis of Computing Systems}, vol.~2, no.~3, pp. 1--41, 2018.

\bibitem{cerf1974protocol}
V.~Cerf and R.~Kahn, ``A protocol for packet network intercommunication,'' \emph{IEEE Transactions on communications}, vol.~22, no.~5, pp. 637--648, 1974.

\bibitem{dataset_criteo_kaggle}
\BIBentryALTinterwordspacing
{Crtieo AI Lab}, ``Kaggle display advertising dataset,'' 2014, accessed: 2014-06-25. [Online]. Available: \url{https://go.criteo.net/criteo-research-kaggle-display-advertising-challenge-dataset.tar.gz}
\BIBentrySTDinterwordspacing

\bibitem{gurobi_optimizer}
\BIBentryALTinterwordspacing
{Gurobi Optimization}, ``Gurobi optimizer reference manual,'' 2024. [Online]. Available: \url{https://docs.gurobi.com/projects/optimizer/en/current/}
\BIBentrySTDinterwordspacing

\bibitem{hrinchuk2019tensorized}
O.~Hrinchuk, V.~Khrulkov, L.~Mirvakhabova, E.~Orlova, and I.~Oseledets, ``Tensorized embedding layers for efficient model compression,'' \emph{arXiv preprint arXiv:1901.10787}, 2019.

\bibitem{kwon2022training}
Y.~Kwon and M.~Rhu, ``Training personalized recommendation systems from (gpu) scratch: Look forward not backwards,'' in \emph{Proceedings of the 49th Annual International Symposium on Computer Architecture}, 2022, pp. 860--873.

\bibitem{zha2022autoshard}
D.~Zha, L.~Feng, B.~Bhushanam, D.~Choudhary, J.~Nie, Y.~Tian, J.~Chae, Y.~Ma, A.~Kejariwal, and X.~Hu, ``Autoshard: Automated embedding table sharding for recommender systems,'' in \emph{Proceedings of the 28th ACM SIGKDD Conference on Knowledge Discovery and Data Mining}, 2022, pp. 4461--4471.

\bibitem{wang2022rec}
Z.~Wang, Y.~Wang, B.~Feng, D.~Mudigere, B.~Muthiah, and Y.~Ding, ``El-rec: Efficient large-scale recommendation model training via tensor-train embedding table,'' in \emph{SC22: International Conference for High Performance Computing, Networking, Storage and Analysis}.\hskip 1em plus 0.5em minus 0.4em\relax IEEE, 2022, pp. 1--14.

\end{thebibliography}

\newpage

\begin{IEEEbiography}[{\includegraphics[width=1in,height=1.25in,clip,keepaspectratio]{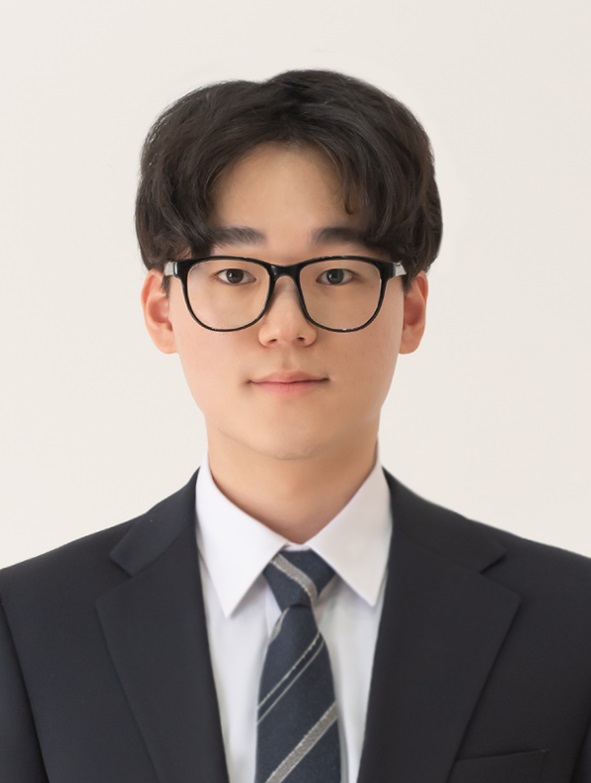}}]{Jinho Yang} Jinho Yang (Graduate Student Member, IEEE) received the B.S. degree in electrical engineering from Hanyang University, Seoul, South Korea, in 2022, and the M.S. degree in electrical engineering from Korea Advanced Institute of Science and Technology (KAIST), Daejeon, South Korea, in 2025, where he is currently pursuing the Ph.D. degree.

His research interests include a hardware accelerator for machine learning, hardware-software co-design, and a near-memory processing system.
\end{IEEEbiography}

\begin{IEEEbiography}[{\includegraphics[width=1in,height=1.25in,clip,keepaspectratio]{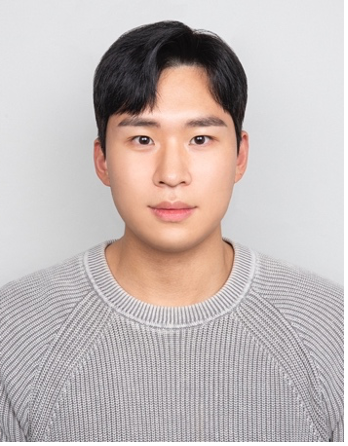}}]{Ji-Hoon Kim} Ji-Hoon Kim (Graduate Student Member, IEEE) received the B.S. degree in electrical engineering from Kyung-Hee University, Suwon, South Korea, in 2017 and the M.S. and Ph.D. degree in electrical engineering from the Korea Advanced Institute of Science and Technology (KAIST), Daejeon, South Korea, in 2019 and 2024, respectively.

His research interests include AI/ML hardware accelerator (ASIC \& FPGA) design, efficient AI/ML system design, energy-efficient processing-in/near-memory architecture, and hardware/software codesign for DNN processing.
\end{IEEEbiography}

\begin{IEEEbiography}[{\includegraphics[width=1in,height=1.25in,clip,keepaspectratio]{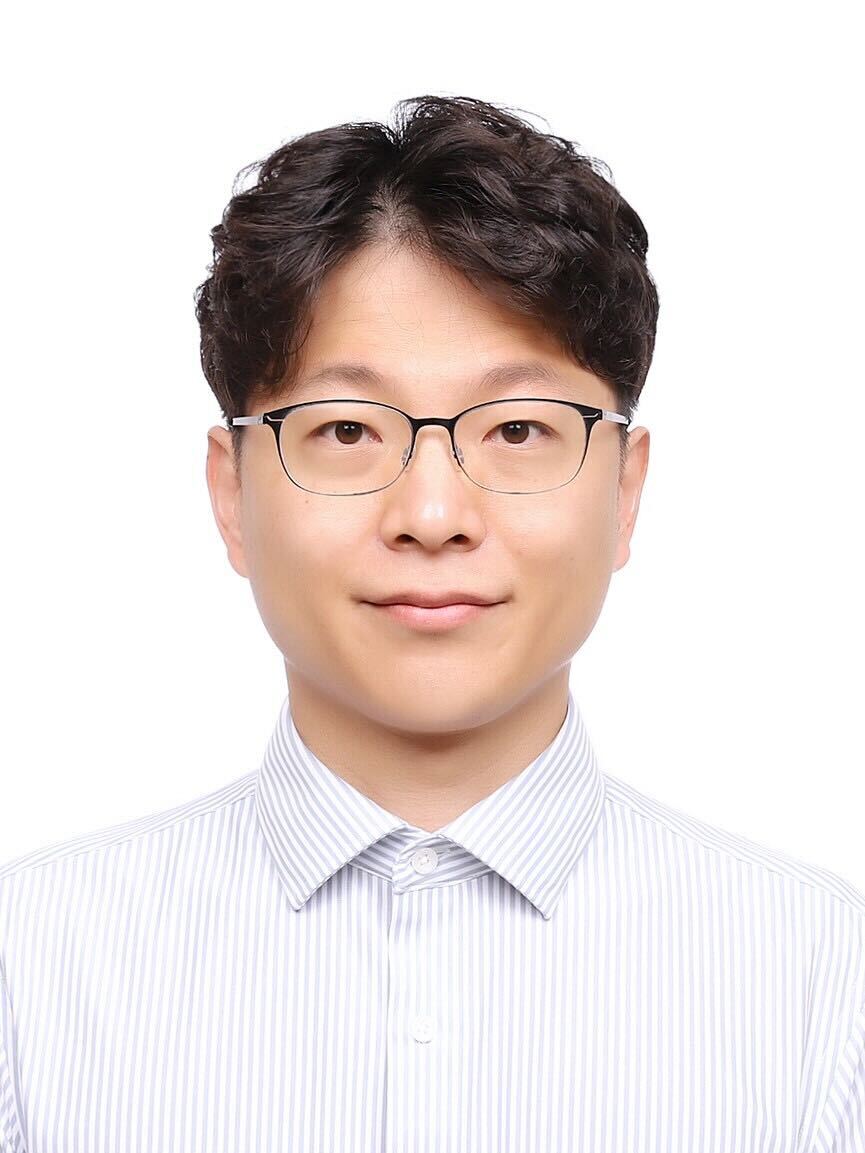}}]{Joo-Young Kim} Joo-Young Kim (Senior Member, IEEE) received the B.S., M.S., and Ph. D degrees in electrical engineering from Korea Advanced Institute of Science and Technology (KAIST), Daejeon, South Korea, in 2005, 2007, and 2010, respectively. He is currently an Associate Professor in the School of Electrical Engineering, KAIST, and the Director of the AI Semiconductor Systems Research Center, KAIST. His research interests span various aspects of hardware design, including chip design, computer architecture, domain-specific accelerators, and hardware/software co-design. Before joining KAIST, he was a Hardware Engineering Leader at Microsoft Azure, Redmond, WA, USA, working on hardware acceleration for cloud services such as machine learning, data storage, and networking.

He founded an AI fabless startup, HyperAccel, in Jan 2023 to build innovative AI processors/solutions for large-language-model(LLM)-based generative AI, making it sustainable for everyone.
\end{IEEEbiography}

\vfill

\end{document}